\documentclass[useAMS,usenatbib]{aa}  

\usepackage{txfonts}
\usepackage{graphicx}
\usepackage{amsmath}
\usepackage{tabularx}
\usepackage{inputenc}
\usepackage{supertabular}
\usepackage{ltxtable}
\usepackage{subfig}
\usepackage{multirow}

\begin{document}

   \title{HADES RV Programme with HARPS-N at TNG}

   \subtitle{V. A super-Earth on the inner edge of the habitable zone of the nearby M-dwarf GJ 625 \thanks{Based on: observations made with the Italian Telescopio Nazionale Galileo (TNG), operated on the island of La Palma by the INAF - Fundación Galileo Galilei at the Roche de Los Muchachos Observatory of the Instituto  de  Astrofísica  de  Canarias (IAC); photometric observations made with the robotic telescope APT2 (within the EXORAP program) located at Serra La Nave on Mt. Etna.; lucky imaging observations made with the Telescopio Carlos S\'anchez operated on the island of Tenerife by the Instituto de Astrof\'isica de Canarias in the Spanish Observatorio del Teide.}}

      \author{A. Su\'{a}rez Mascare\~{n}o\inst{1,2,3} \and
         J.~I. Gonz\'{a}lez Hern\'{a}ndez   \inst{1,3} \and
         R. Rebolo  \inst{1,3,4}   \and
         S. Velasco\inst{1,3} \and 
         B. Toledo-Padr\'{o}n \inst{1,3} \and
         L. Affer\inst{5} \and
         M. Perger\inst{6} \and
         G. Micela\inst{5} \and
         I. Ribas\inst{6} \and
         J. Maldonado\inst{5} \and
         G. Leto\inst{7}\and
         R. Zanmar Sanchez\inst{7} \and
         G. Scandariato\inst{7} \and
         M. Damasso\inst{8} \and
         A. Sozzetti\inst{8} \and
         M. Esposito \inst{9} \and
         E. Covino \inst{9} \and
         A. Maggio \inst{5} \and
         A. F. Lanza \inst{7} \and
         S. Desidera \inst{10} \and
         A. Rosich\inst{6} \and   
         A. Bignamini \inst{11} \and
         R. Claudi \inst{10} \and
         S. Benatti\inst{10} \and
         F. Borsa \inst{12}\and
         M. Pedani\inst{13} \and
         E. Molinari\inst{13} \and
         J. C. Morales \inst{6} \and
         E. Herrero \inst{6} \and
         M. Lafarga \inst{6}
}         
    \institute{Instituto de Astrof\'{i}sica de Canarias, E-38205 La Laguna, Tenerife, Spain\\
                 \email{jonay@iac.es}          \and
             Observatoire Astronomique de l'Université de Genève, Versoix, Switzerland  \\
                  \email{Alejandro.SuarezMascareno@unige.ch}          \and
             Universidad de La Laguna, Dpto. Astrof\'{i}sica, E-38206 La Laguna, Tenerife, Spain \and
             Consejo Superior de Investigaciones Cient{\'\i}ficas, Spain \and
             INAF - Osservatorio Astronomico di Palermo, Piazza del Parlamento 1, 90134 Palermo, Italy \and
             Institut de Ciènces de l'Espai (CSIC-IEEC), Campus UAB, Carrer de Can Magrans s/n, 08193 Cerdanyola del Vallés, Spain \and
 		     INAF - Osservatorio Astrofisico di Catania, via S. Sofia 78, 95123 Catania, Italy \and            
             INAF - Osservatorio Astrofisico di Torino, via Osservatorio 20, 10025 Pino Torinese, Italy \and
             INAF - Osservatorio Astronomico di Capodimonte, Via Moiariello, 16, 80131 - NAPOLI \and
             INAF - Osservatorio Astronomico di Padova, Vicolo dell’Osservatorio 5, 35122, Padova, Italy \and
             INAF - Osservatorio Astronomico di Trieste, via Tiepolo 11,34143, Trieste, Italy \and             
             INAF - Osservatorio Astronomico di Brera, Via E. Bianchi 46,I-23807 Merate (LC), Italy \and
             Fundación Galileo Galilei - INAF, Rambla José Ana Fernandez Pérez 7, E-38712 Breña Baja, TF - Spain
             }

   \date{Revised April-2017}

 
  \abstract
   {We report the discovery of a super-Earth orbiting at the inner edge of the habitable zone of the star GJ 625 based on the analysis of the radial-velocity (RV) time series from the HARPS-N spectrograph, consisting in 151 HARPS-N measurements taken over 3.5 yr. GJ 625 b is a planet with a minimum mass M sin $i$ of 2.82 $\pm$ 0.51 M$_{\oplus}$ with an orbital period of 14.628 $\pm$ 0.013 days at a distance of 0.078 AU of its parent star. The host star is the quiet M2 V star GJ 625, located at 6.5 pc from the Sun. We find the presence of a second radial velocity signal in the range 74-85 days that we relate to stellar rotation after analysing the time series of Ca II H\&K and H${\alpha}$ spectroscopic indicators, the variations of the FWHM of the CCF and and the APT2 photometric light curves. We find no evidence linking the short period radial velocity signal to any activity proxy.
}
   
   {}

   \keywords{
              Planetary Systems --- Techniques: radial velocity --- Stars: activity --- Stars: chromospheres --- Stars: rotation --- Stars: magnetic cycle --- starspots --- Stars: individual (GJ 625)
 }

   \maketitle
%

\section{Introduction}

The detection rate of potentially habitable Earth-like planets around M-dwarfs is rapidly increasing  \citep{Wright2016, AngladaEscude2016, Jehin2016, Gillon2017,AstudilloDefru2017}. Since years ago it has been clear that M-dwarfs are a shortcut to find earth-like planets and therefore several surveys have attempted to take advantage of their low masses and closer habitable zones \citep{Quirrenbach2012, Bonfils2013, Howard2014, Irwin2015, Berta-Thompson2015, Affer2016, Masca2017, Perger2017b}. M-dwarfs have also proven to be difficult targets because of their stellar activity. The signals induced by the stellar rotation can easily mimic those of planetary origin \citep{Queloz2001, Bonfils2007,Boisse2011, Robertson2013,Masca2015,Newton2016,Vanderburg2016, Masca2017,Masca2017b}. For the case of M-dwarfs these signals tend to be comparable to those of rocky planets close to the habitable zone of their stars \citep{Howard2014, Robertson2013,Masca2015,Newton2016,Masca2017b}. The interpretation of the radial velocity curves of M-dwarfs is usually complicated even in the quietest of stars. Low mass stars are the most common type of stars, offering valuable complementary information on the formation mechanisms of planetary systems. For instance giant planets at close orbits are known to be rare around M dwarfs \citep{Endl2006}, while low mass rocky planets appear to be more frequent \citep{Bonfils2013, Dressing2013, Dressing2015}.

Even with the large amount of confirmed exoplanets \citep{Howard2009, Mayor2011, Howard2012} the number of rocky planets remains comparably small with only about a hundred known around M-dwarfs. This is mostly because most surveys -- including the Kepler survey -- have focused primarily in solar type and K-type stars. In the recent years many planetary systems have been reported hosting Neptune mass planets and super-Earths \citep{Udry2007, Delfosse2013,Howard2014,AstudilloDefru2015}, and a rapidly increasing amount of Earth-mass planets \citep{Mayor2009,BertaThompson2015, Wright2016, Affer2016, AngladaEscude2016, Jehin2016, Masca2017, Gillon2017}. However the frequency of Earth-sized planets around M-dwarfs is still not well established. Several studies have attempted to quantify the abundance of rocky planets in close orbits and in the habitable zones of M-dwarfs, but the uncertainties are still large, making it important to continue adding new planets to the sample \citep{Bonfils2013, Gaidos2013, Kopparapu2013}.

We report the discovery of a super-Earth orbiting the nearby star GJ 625 on the inner edge of its habitable zone. The discovery is part of the HADES (HArps-n red Dwarf Exoplanet Survey) radial velocity (RV) program with HARPS-N at the Telescopio Nazionale Galileo in La Palma (Spain). The HADES RV program is the result of a collaborative effort between the Italian Global Architecture of Planetary Systems (GAPS,~\citet{Covino2013,Poretti2016}) Consortium, The Institut de Ci\`{e}ncies de l'Espai de Catalunya (ICE), and the Instituto de Astrof\'{i}sica de Canarias (IAC). The HADES team has previously discovered two super-Earth exoplanets of minimum masses of $\sim$ 6.3 and $\sim$ 2.5 M$_{\oplus}$ orbiting the early-type M-dwarf GJ 3998 \citep{Affer2016}.

GJ 625 is a bright (V = 10.17 mag) low activity ($\log_{10}(R'_\textrm{HK})$ $\sim$ -- 5.5) M-dwarf located at a distance of 6.5 pc from the Sun \citep{vanLeeuwen2007,Gaia2016}. Table~\ref{parameters} shows the stellar parameters. Its low activity combined with its long rotation period, of more than 70 days, makes it a very interesting candidate to search for rocky planets. 

\begin {table}
\begin{center}
\caption { Stellar parameters of GJ 625 \label{tab:parameters}}
    \begin{tabular}{ l  l  l l l l l l l l l l } \hline
Parameter  & GJ 625 & Ref.\\ \hline
RA (J2000) & 16:25:25.36 & 1\\
DEC (J2000) & +54:18:12.19 & 1\\
$\mu_{\alpha}$ ($mas$ yr$^{-1}$) & 432.13 & 1\\
$\mu_{\delta}$ ($mas$ yr$^{-1}$) & --171.48 & 1\\
Distance [pc] & 6.49 & 1 \\
$m_{B}$ & 11.80 $\pm$ 0.16 & 2\\
$m_{V}$			& 10.17 $\pm$ 0.04 & 2 \\	
Spectral Type & M2 & 3\\
T$_{\rm eff}$ [K] & 3499 $\pm$ 68 & 3\\
$[Fe/H]$ & -0.38 $\pm$ 0.09 & 3\\
$M_\star$~$[M_{\odot}]$ & 0.30 $\pm$ 0.07 & 3\\
$R_\star$~$[R_{\odot}]$ & 0.31 $\pm$ 0.06 & 3\\
log $g$ (cgs) &4.94 $\pm$ 0.06 & 3\\
log($L_{\star}/L_{\odot}$) & --1.894 $\pm$ 0.170 & 3 \\
$\log_{10}(R'_\textrm{HK})$ &  --5.5 $\pm$ 0.2 & 0\\
$P_{\rm rot}$ (days) & 77.8 $\pm$ 5.5 & 0\\
$v \sin i$ (km s$^{-1}$) & $\textless$ 0.2$^{*}$ &  3\\
Secular acc. (m s$^{-1}$ yr$^{-1}$) & 0.03 & 4 \\
 \hline
\label{parameters}
\end{tabular}  
\end{center}
\textbf{References:} 0 - This work, 1 - \citet{Gaia2016}, 2 - \citet{Hog2000}, 3 -\citet{Maldonado2017}, 4 - Calculated following \citet{Montet2014}.\\
$^{*}$ Calculated using the radius estimated by \citet{Maldonado2017} and our period determination. \\
\end {table}

\section{Data \& Observations}

\subsection{Spectroscopy}
The star GJ 625 is part of the HADES RV program \citep{Affer2016} and has been extensively monitored since 2013. We have used 151 HARPS-N spectra taken over 3.5 yr. HARPS-N \citep{Cosentino2012} is a fibre-fed high resolution echelle spectrograph installed at the Telescopio Nazionale Galileo in the Roque de los Muchachos Observatory (Spain). The instrument has a resolving power greater than $R\sim 115\,000$ over a spectral range from $\sim$380 to $\sim$690 nm and has been designed to attain very high long-term radial-velocity precision. It is contained in a vacuum vessel to avoid spectral drifts due to temperature and air pressure variations, thus  ensuring its stability. HARPS-N is equipped with its own pipeline providing extracted and wavelength-calibrated spectra, as well as RV measurements and other data products such as cross-correlation functions and their bisector profiles. 

Most of the observations were carried out using the Fabry Perot (FP) as simultaneous calibration. The FP offers the possibility of monitoring the instrumental drift with a precision of 10 cms$^{-1}$ without the risk of contamination of the stellar spectra by the ThAr saturated lines \citep{Wildi2010}. While this is not usually a problem in G and K stars, the small amount of light collected in the blue part of the spectra of M-dwarfs might compromise the quality of the measurement of the Ca II H\&K flux.  The FP allows a precision of $\sim$ 1 ms$^{-1}$ in the determination of the radial velocities of the spectra with highest signal to noise ratios while assuring the quality of the spectroscopic indicators even in those spectra with low signal to noise ratios. Those observations taken without the FP were taken without simultaneous lamp reference. 

\subsection{EXORAP photometry}

Along with the HARPS-N spectra we have 2.6 yr of B, V, R, and I-band photometric data collected at the INAF-Catania Astrophysical  Observatory with an 80 cm f/8 \textit{Ritchey-Chretien} robotic telescope (APT2) located at Serra la Nave on the Mt. Etna . The data is reduced by overscan, bias, dark subtraction and flat fielding with IRAF \footnote{IRAF is distributed by the National Optical Astronomy Observatories, which are operated by the Association of Universities for Research in Astronomy, Inc., under cooperative agreement with the National Science Foundation.} procedures and visually inspected to check the quality (see \citet{Affer2016} for details). Errors in the individual measurements are the quadratic sum of the intrinsic noise (photon noise and sky noise) and the RMS of the ensemble stars used for the differential photometry.

\subsection{FastCam Lucky Imaging Observations}

On June 6th 2016, we collected 50,000 individual frames of GJ 625 in the I band using the lucky imaging FastCam instrument
\citep{Oscoz2008} at the 1.5m Carlos S\'anchez Telescope in the Observatorio del Teide, Tenerife, with 30 ms exposure time for each
frame. FastCam is an optical imager with a low noise EMCCD camera which allows to obtain speckle-featuring not saturated images at a fast frame rate \citep{Labadie2011}.

\begin{figure}
\centering
	\fbox{\includegraphics[width=8.5cm]{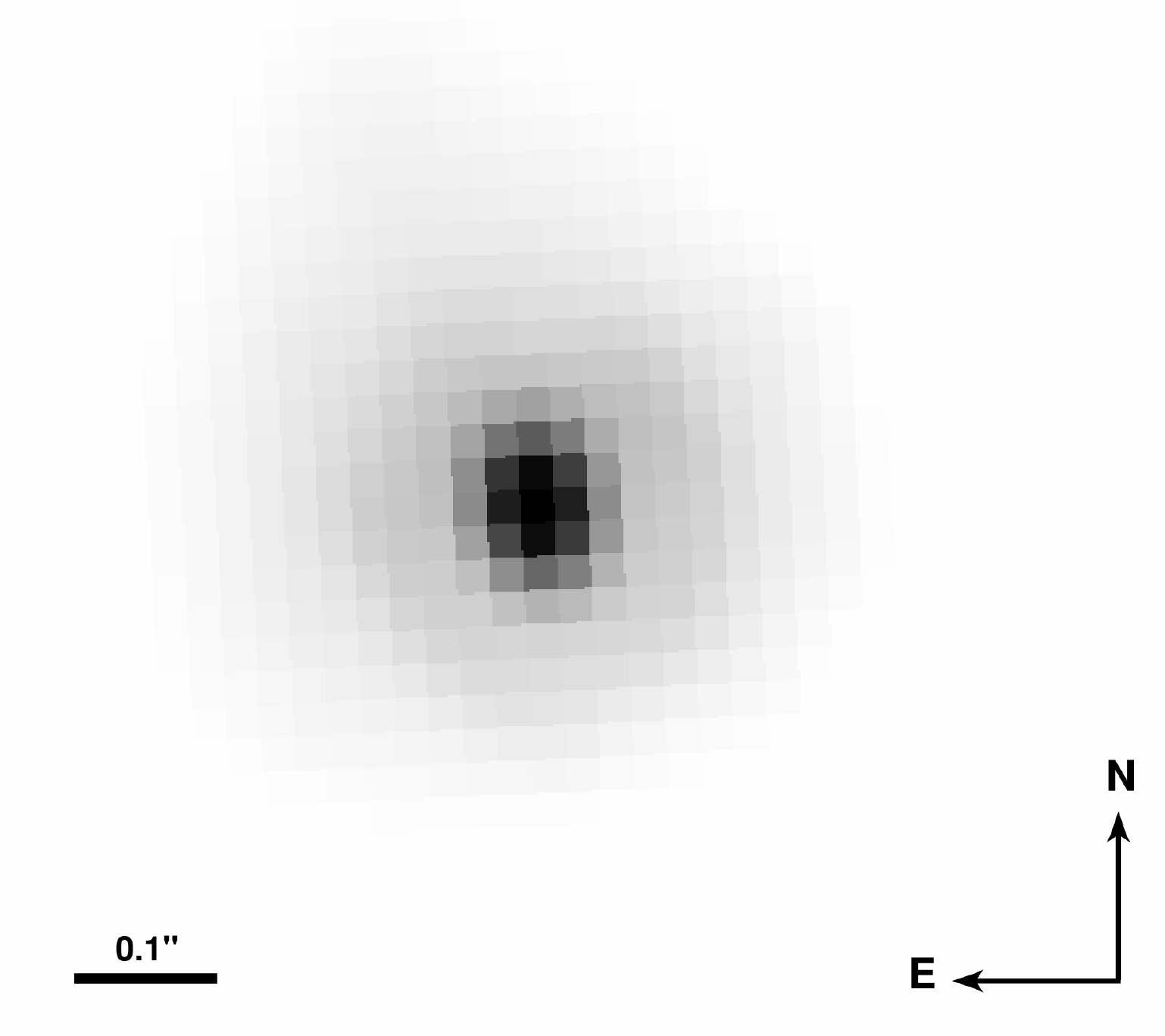}}
	\caption{Diffraction-limited image of GJ 625 after lucky imaging processing with a selection of the 30\% best individual TCS/FastCam frames.}
	\label{lucky}
 \end{figure}

In order to construct a high resolution, diffraction limited, long-exposure image, the individual frames were bias subtracted, aligned and co-added using our own Lucky Imaging  algorithm \citep{Velasco2016}. Figure \ref{lucky} presents the high
resolution image constructed by co-addition of the  best percentage of the images using lucky imaging and shift-and-add algorithms. Due to the atmospheric conditions of the night, the selection of 30\% of the individual frames was found to be the best solution to  produce a deep and diffraction limited image of the target, resulting in a total integration time of 450 seconds.  The combined image achieved $\Delta m_I=4.5 - 5.0$ at $1 \farcs$ We find no bright contaminant star in the diffraction limited image.

\section{Determination of Stellar Activity Indicators and Radial Velocities}

\subsection{Activity Indicators}
For the activity analysis we use the extracted order-by-order wavelength-calibrated spectra produced by the HARPS-N pipeline. For a given star, the change in atmospheric transparency from day to day causes variations in the  flux distribution of the recorded spectra that are particularly relevant in the blue, where we intend to measure Ca II lines. In order to minimize  the effects related to these atmospheric changes  we create a spectral template for each star by de-blazing and co-adding every available spectrum and use the co-added spectrum to correct the order-by-order fluxes of the individual ones. We also   correct each spectrum for the Earth's barycentric radial velocity and the radial velocity of the star using the measurements given by the standard pipeline and re-binned the spectra into a wavelength-constant step. Using this  HARPS-N dataset, we expect to have high quality spectroscopic indicators  to monitor tiny stellar activity variations with  high accuracy. 

\subsection*{S$_{MW}$ Index}

\begin{figure}
\centering
	\includegraphics[width=9.0cm]{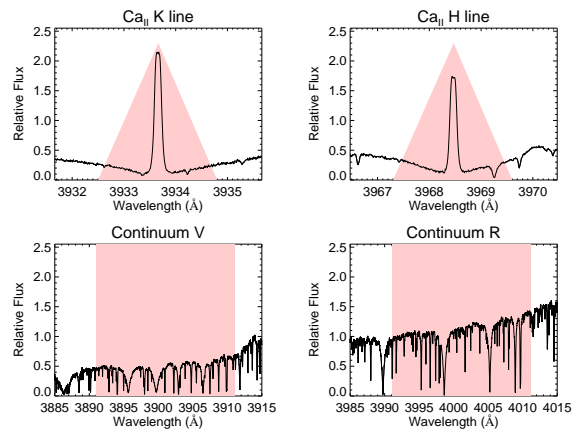}
	\caption{Ca II H\&K filter of the spectrum of the star GJ 625 with the same shape as the Mount Wilson Ca II H\&K passband.}
	\label{Sindex}
\end{figure}

We calculate the Mount Wilson $S$ index and the $\log_{10}(R'_{HK})$ by using the original \citet{Noyes1984} procedure, following \citet{Lovis2011} and \citet{Masca2015, Masca2017}. We define two triangular-shaped passbands with  full width half maximum (FWHM) of 1.09~{\AA} centred at 3968.470~{\AA} and 3933.664~{\AA} for the Ca II H\&K line cores, and for the continuum we use two 20~{\AA} wide bands centred at 3901.070~{\AA} (V) and 4001.070~{\AA}(R), as shown in figure~\ref{Sindex}.

Then the S-index is defined as 

\begin{equation}
   S=\alpha {{\tilde{N}_{H}+\tilde{N}_{K}}\over{\tilde{N}_{R}+\tilde{N}_{V}}} + \beta,
\end{equation}
\noindent where $\tilde{N}_{H},\tilde{N}_{K},\tilde{N}_{R}$ and $\tilde{N}_{V}$ are the mean fluxes per wavelength unit in each passband,  while $\alpha$ and $\beta$ are calibration constants fixed as $\alpha = 1.111$ and $\beta = 0.0153$ . The S index (S$_{MW}$) serves as a measurement of the Ca II H\&K core flux normalized to the neighbour continuum. As a normalized index to compare it to other stars we compute the $\log_{10}(R'_\textrm{HK})$ following \citet{Masca2015}.

\subsection*{H$\alpha$ Index}

\begin{figure}
	\includegraphics[width=9.0cm]{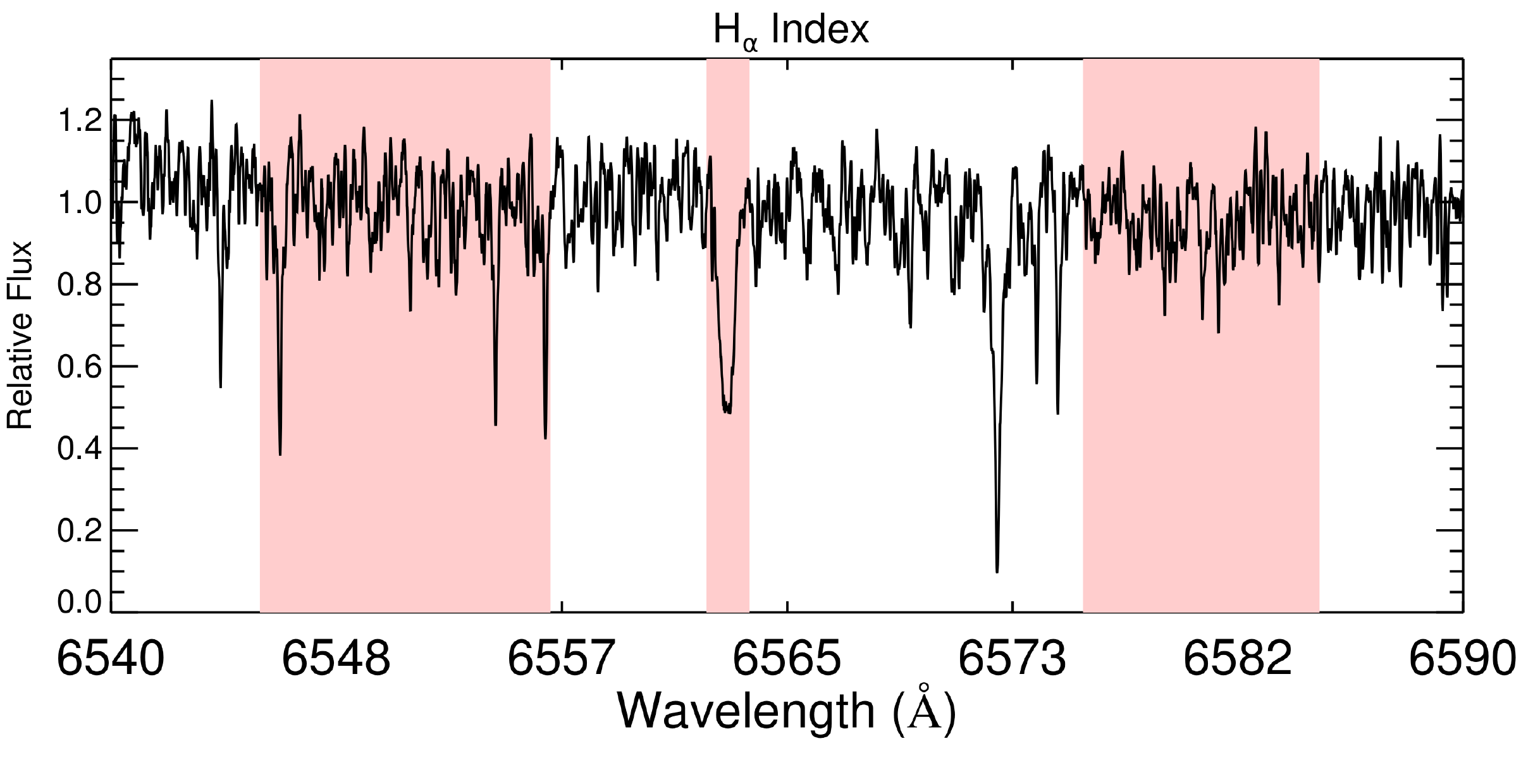}
	\caption{Spectrum of the M-type star GJ 625 showing the H$\alpha$ region of the spectrum. The light-red shaded regions show the filter passband and continuum bands.}
	\label{halpha}
\end{figure}

We also use the H$\alpha$ index, with a simpler passband following \citet{GomesdaSilva2011} and \citet{Masca2017}. It consists of a rectangular bandpass with a width of 1.6~{\AA} and centred at 6562.808~{\AA} (core), and two continuum bands of 10.75~{\AA} and a 8.75~{\AA} wide centred at 6550.87~{\AA} (L) and 6580.31~{\AA} (R), respectively, as seen in Figure~\ref{halpha}. 

Thus, the H$\alpha$ index is defined as

\begin{equation}
   H\alpha_{\rm Index}={{\tilde{H}\alpha_{\rm core} }\over{\tilde{H}\alpha_{L} +\tilde{H}\alpha_{R}}}.
\end{equation}
 
Where $\tilde{H}\alpha_{\rm core}, \tilde{H}\alpha_{L}$ and $\tilde{H}\alpha_{R}$ are the mean fluxes per wavelength unit in each of the previously defined passbands.

\subsection{Radial velocities}

The radial-velocity measurement in the HARPS-N standard pipeline is determined by a Gaussian fit of the cross correlation function (CCF) of the spectrum with a synthetic stellar template \citep{Baranne1996, PepeMayor2000} that consists on a series of delta functions at the positions of isolated stellar lines at wavelength longer than 4400~{\AA}. In the case of M-dwarfs, due to the huge number of line blends, the cross correlation function is not Gaussian, resulting in a less precise Gaussian fit which might cause distortions in the radial-velocity measurements and decrease the sensibility in the measurement of the variations in the full width half maximum (FWHM) of the CCF. To deal with this issue we tried two different approaches as in \citet{Masca2017}. 

The first one consisted in using a slightly more complex model for the CCF fitting, a Gaussian function plus a second order polynomial using only the central region of the CCF function. We use a 15 km s$^{-1}$ window centred at the minimum of the CCF. This configuration provides the best stability of the measurements. Along  with the measurements of the radial velocity, we obtain the FWHM of the cross correlation function which we also use to track variations in the activity level of the star. A second approach to the problem was to recompute the radial velocities using the TERRA pipeline \citep{AngladaEscude2012b}, which is a template matching algorithm with a high signal to noise stellar spectral template. Every spectrum is corrected for both barycentric and stellar radial velocity to align it to the frame of the solar system barycenter. The radial velocities are computed by minimizing the $\chi^2$ of the residuals between the observed spectra and shifted versions of the stellar template, with all the elements contaminated by telluric lines masked. In this case we are using the Doppler information at wavelength longer than 4550 ~{\AA}. All radial-velocity measurements are corrected from the secular acceleration of the star.

For the bisector span measurement we rely on the pipeline results, as it does not depend on the fit but on the CCF itself. The bisector has been a standard activity diagnostic tool for solar type stars since more than 10 years ago . Unfortunately its behaviour in slow rotating stars is not as informative as it is for fast rotators \citep{Saar1997, Bonfils2007}. We report the measurements of the bisector span (BIS) for each radial-velocity measurement, but we do not find any meaningful information in its analysis.

\subsection{Quality Control of the Data}

As the sampling rate of our data is not  well suited for modelling fast events, such as flares, and their effect in the radial velocity is not well understood, we identify and reject points likely affected by flares by searching for an abnormal behaviour of the activity indicators \citep{Reiners2009}. The process rejected 11 spectra that correspond to flare events of the star with obvious activity enhancement and line distortion. That leaves us with 140 HARPS-N spectroscopic observations taken over 3.3 years with  a typical exposure of 900 s and an average signal to noise ratio of 56 at $5500$~{\AA}. All measurements and their uncertainties are reported in Table~\ref{tab:full_data}.

\section{Analysis}

In order to properly understand the behaviour of the star, our first step is to analyse the different modulations present in the photometric and spectroscopic time-series. 

We search for periodic variability compatible with both stellar rotation and long-term magnetic cycles. We compute the power spectrum using a Generalised Lomb Scargle Periodogram \citep{Zechmeister2009} and if there is any significant periodicity we fit the detected period using sinusoidal model, or a double harmonic sinusoidal model to account for the asymmetry of some signals \citep{BerdyuginaJarvinen2005}, with the MPFIT routine \citep{Markwardt2009}. The model used for the preliminary analysis is chosen based on the reduced $\chi^2$ of the fit. 

The significance of the periodogram peak is evaluated using both the \citet{Cumming2004} modification of the \citet{HorneBaliunas1986} formula to obtain the spectral density thresholds for a desired false alarm probability (FAP) level and bootstrap randomization \citep{Endl2001} of the data. We finally opted for the power spectral density (PSD) levels given by the bootstrap process as those where the most conservative. 

\subsection*{Radial velocities}

\begin{figure}
	\includegraphics[width=9.0cm]{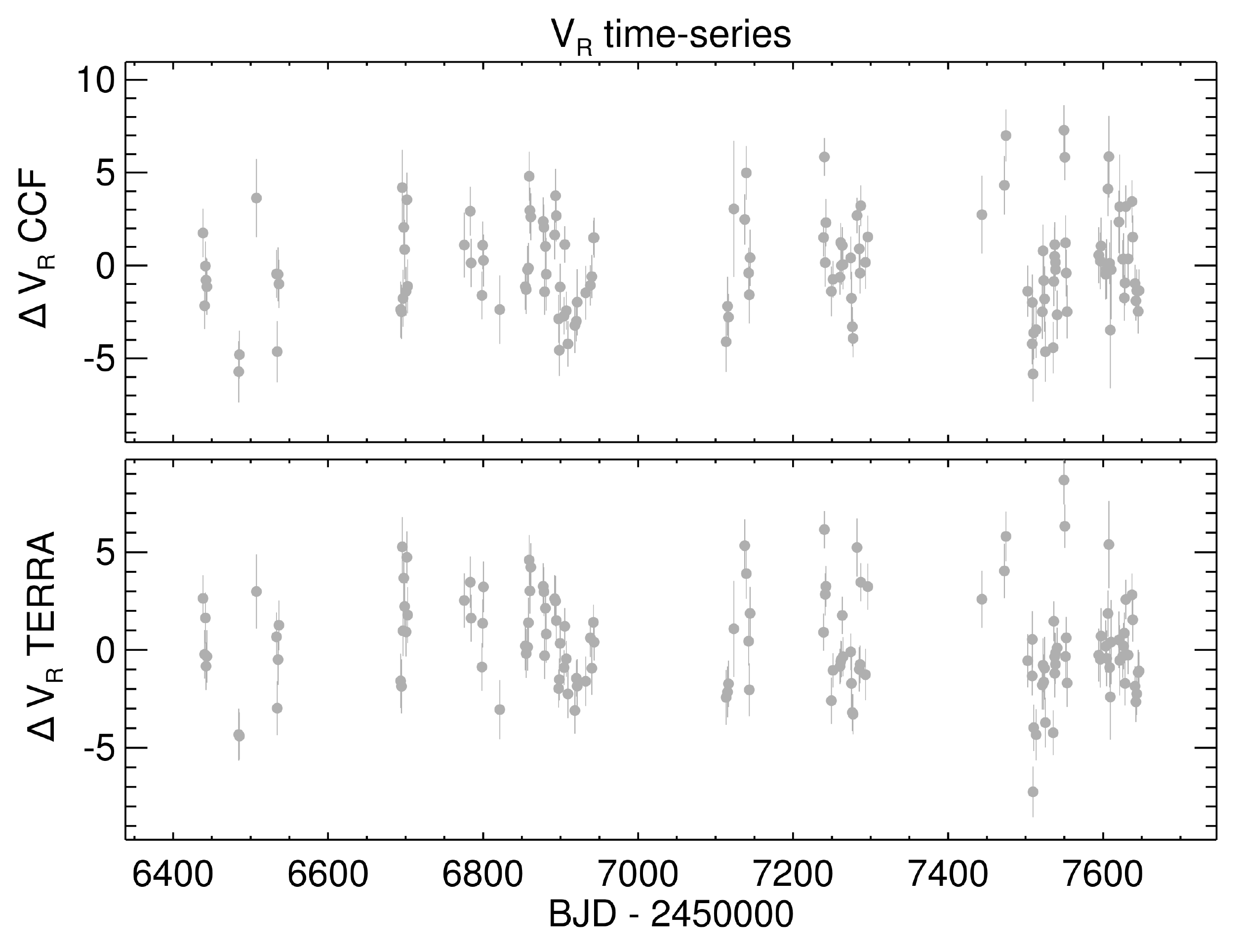}
	\caption{Radial-velocity time series for the CCF RVs (top panel) and the TERRA RVs (bottom panel).}
	\label{RV_Series}
\end{figure}

For the analysis of the radial velocities we use both the measurements given by the CCF and by the TERRA pipeline. We have 140 measurements distributed along 3.3 years. The star shows a mean RV of -12.850 Kms$^{-1}$. The CCF RV measurements show an RMS of 2.56 ms$^{-1}$ with a typical uncertainty of 1.32 m s$^{-1}$ while the TERRA RV measurements show an RMS of 2.57 m s$^{-1}$ with a typical uncertainty of 1.24 m s$^{-1}$. Figure~\ref{RV_Series} shows both time-series of RVs. Both sets show a very similar scatter with the TERRA data having slightly smaller error bars.

\begin{figure}
	\includegraphics[width=9.0cm]{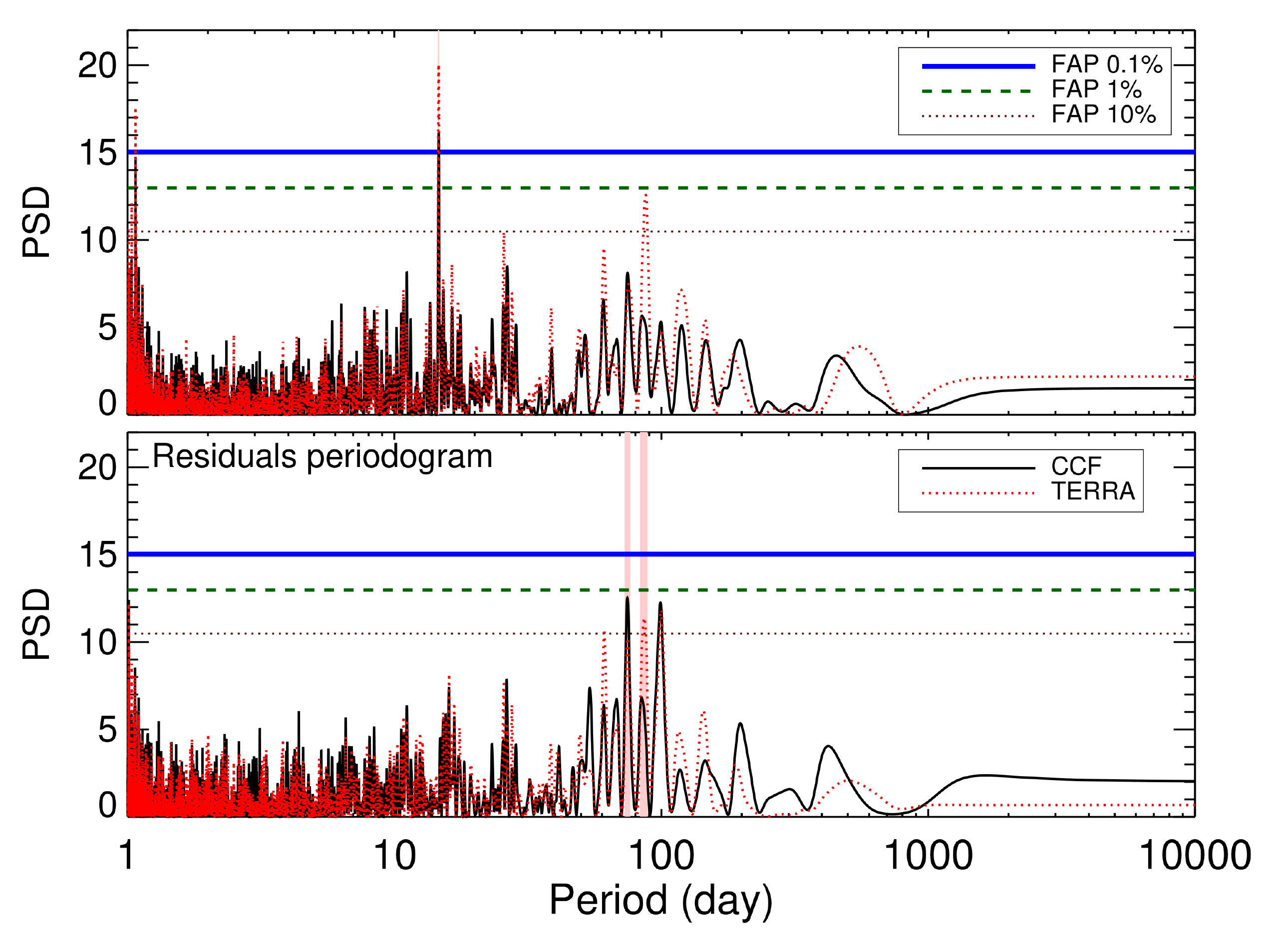}
	\caption{Periodograms of the radial velocity time series. Top panel shows the periodogram of the raw RVs for both the CCF (black line) and TERRA  (red dotted line). Bottom panel shows the periodogram of the residuals for both the CCF (black line) and TERRA (red dotted line). The light-red shaded regions show the detected periodicities.}
	\label{RV_periodograms}
\end{figure}

\begin{figure*}
\begin{minipage}{0.5\textwidth}
        \centering
        \includegraphics[width=9.cm]{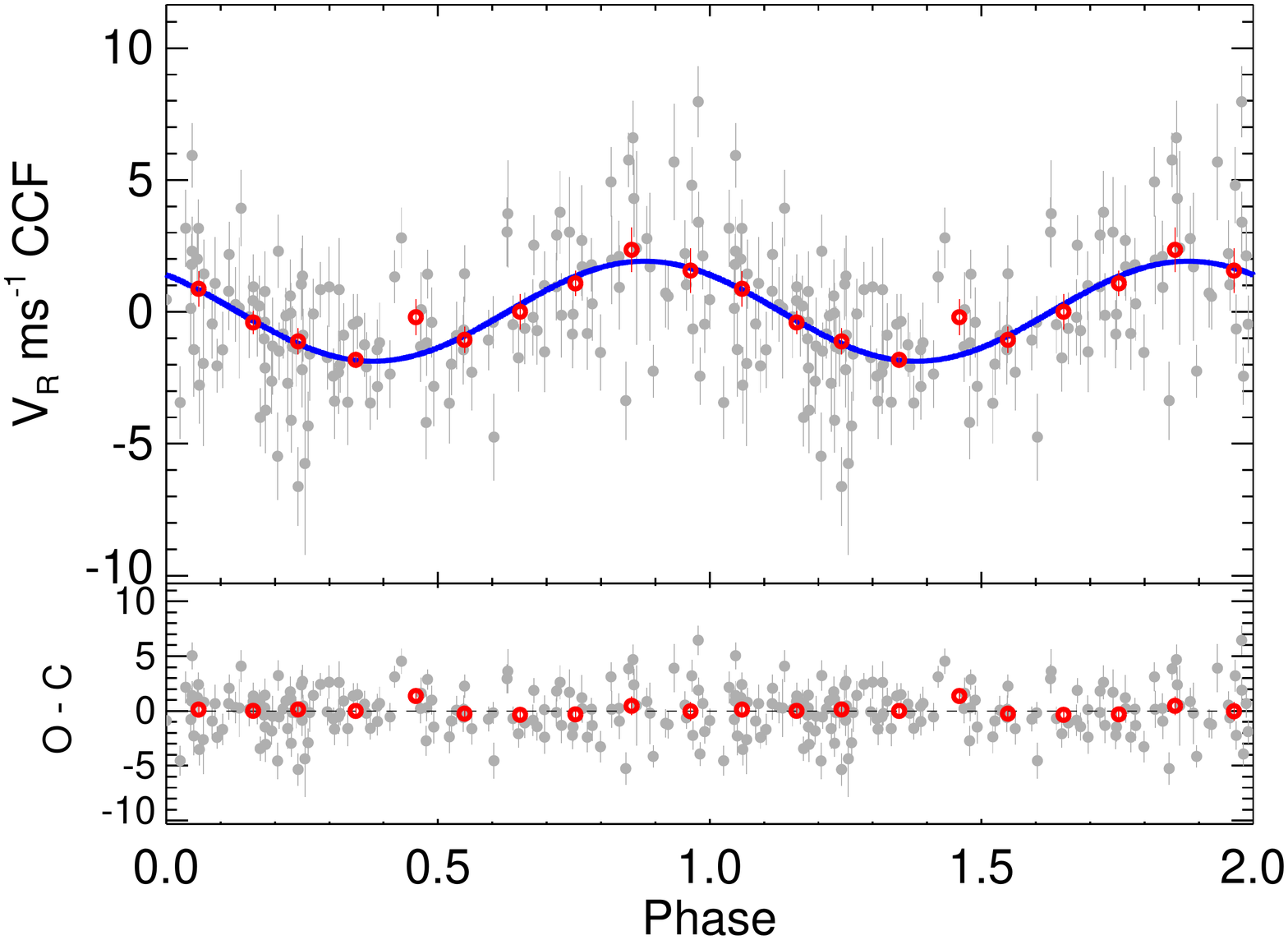}
\end{minipage}%
\begin{minipage}{0.5\textwidth}
        \centering
        \includegraphics[width=9.cm]{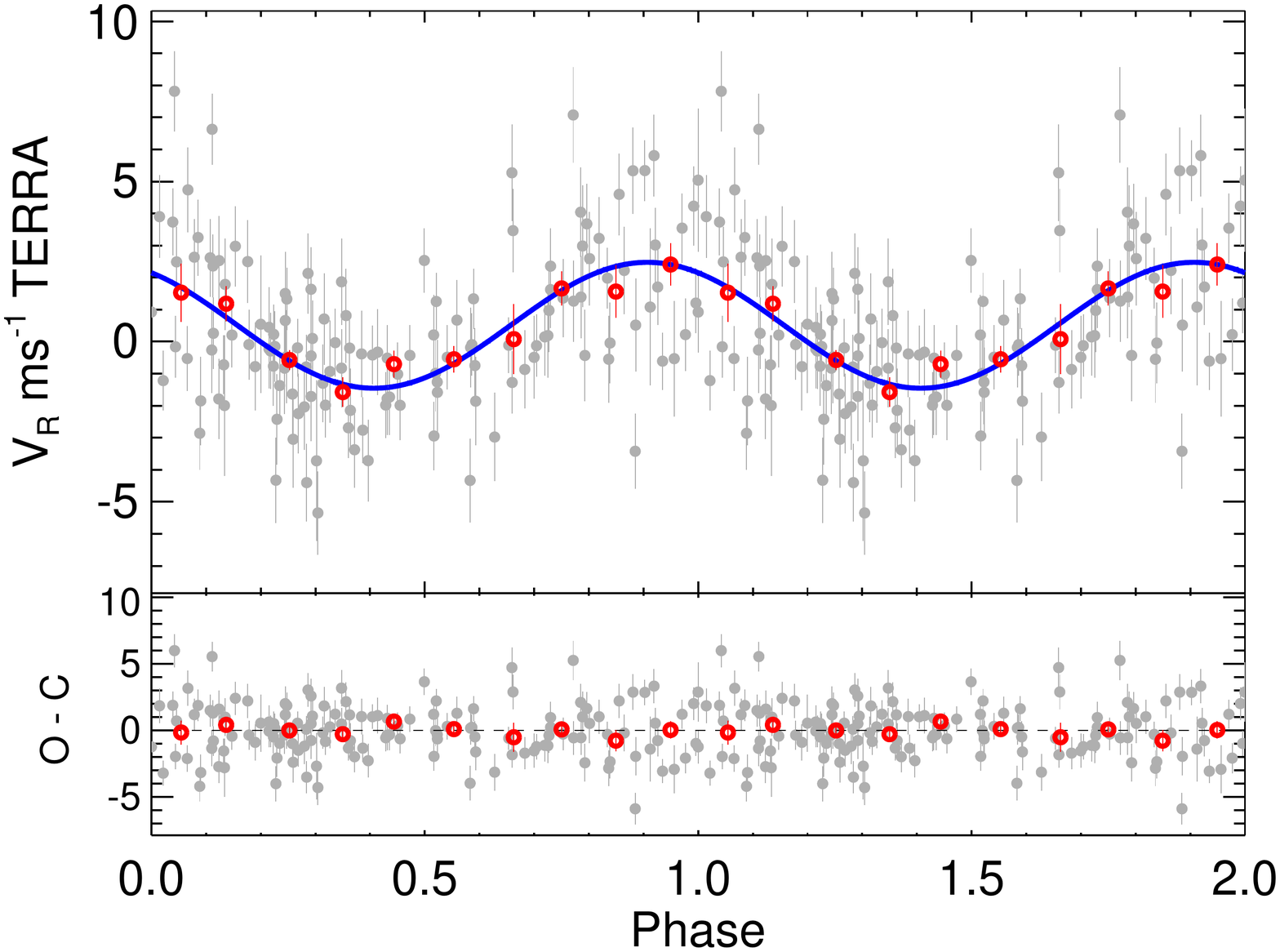}
\end{minipage}%
\caption{Phase folded curve of the raw RV data using the 14.6 days periodicity. Left panel shows the CCF measurements, right panel the TERRA measurements. Red dots are the same points binned in phase with a bin size of 0.1. The error bar of a given bin is estimated using the weighted standard deviation of binned measurements divided by the square root of the number of measurements included in this bin. The blue solid line shows the best fit to the data using a sinusoidal fit.}
\label{RV_pla}
\end{figure*}

\begin{figure*}
\begin{minipage}{0.5\textwidth}
        \centering
        \includegraphics[width=9.cm]{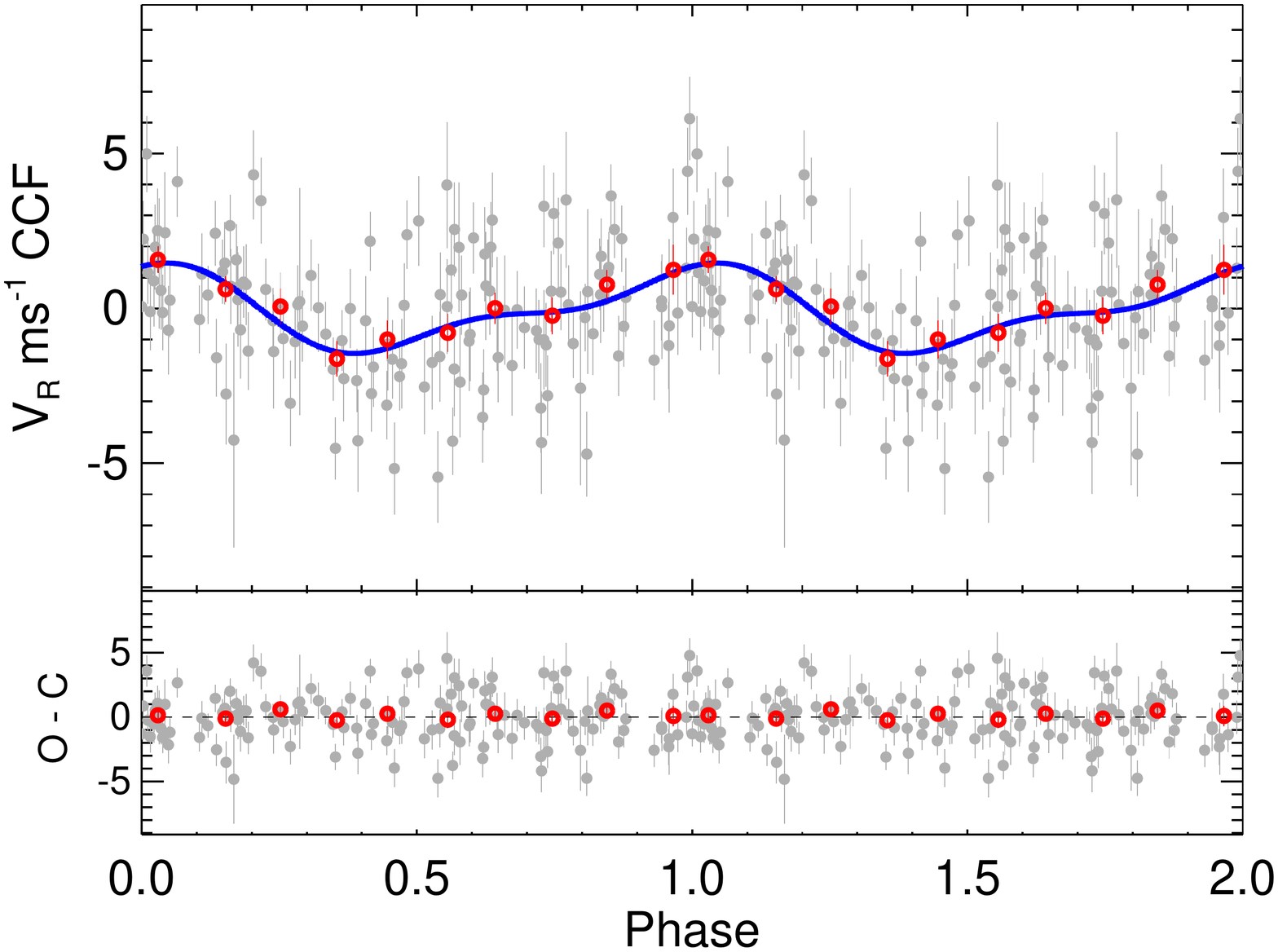}
\end{minipage}%
\begin{minipage}{0.5\textwidth}
        \centering
        \includegraphics[width=9.cm]{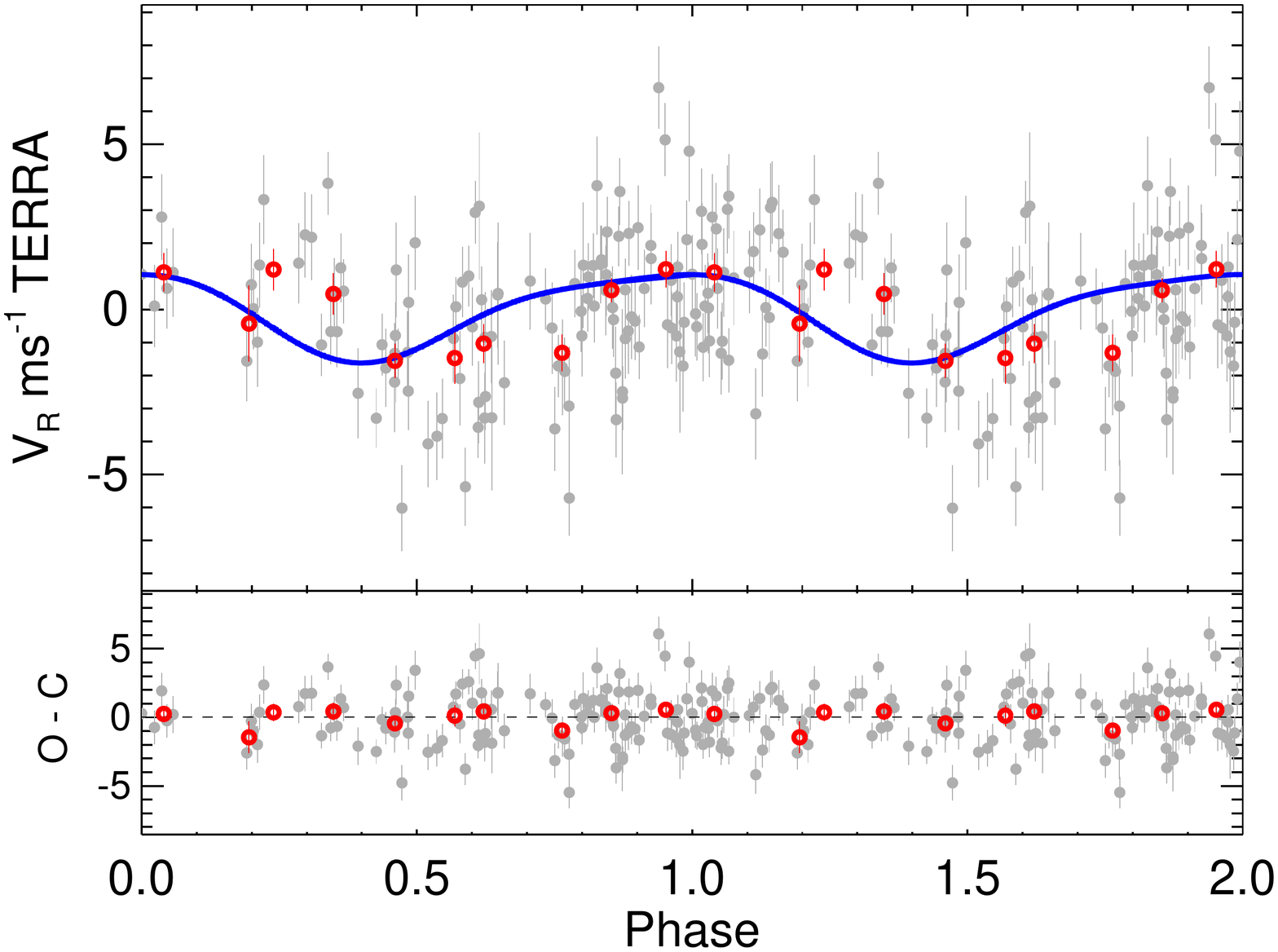}
\end{minipage}%
\caption{Phase folded curve of the RV data using the 74 days periodicity for the CCF data (left panel) and the 86 days periodicity for the TERRA data (right panel) after subtracting the signals shown in Fig.~\ref{RV_pla}. Red dots are the same points binned in phase with a bin size of 0.1. The error bar of a given bin is estimated using the weighted standard deviation of binned measurements divided by the square root of the number of measurements included in this bin. The blue solid line shows the best fit to the data using a double harmonic sinusoidal fit.}
\label{RV_rot}
\end{figure*}

We find a clear dominant signal in the data, present in both series with a FAP <0.1\% (although more significant in the TERRA data). The signal has a period of 14.629 $\pm$ 0.069 days in the CCF series and 14.629 $\pm$ 0.077 days in the TERRA series, a semi amplitude of 1.85 $\pm$ 0.13 ms$^{-1}$ in the CCF data, and a semi amplitude of 1.65 $\pm$ 0.18 ms$^{-1}$ in the TERRA data. Figure~\ref{RV_periodograms} shows the periodograms of both data series combined and Figure~\ref{RV_pla} shows the phase folded fits of both series. When analysing the residuals we see a structure of different signals between $\sim$ 65 to $\sim$ 100 days (see Fig.~\ref{RV_periodograms}). Of those the most significant one is not the same in the CCF and TERRA data. The CCF data shows a 74.7 $\pm$ 1.9 d signal with a semi amplitude of 1.64 $\pm$ 0.18 ms$^{-1}$ while the TERRA data shows a 85.9 $\pm$ 2.8 d signal with a semi amplitude of 1.58 $\pm$ 0.18 ms$^{-1}$. Figure~\ref{RV_rot} shows the phase folded curves for those signals. No more significant signals are found in the RV data. The RMS of the remaining residuals is 2.0 ms$^{-1}$ in both cases. 

\subsection*{Spectroscopic indicators}

\begin{figure}
	\includegraphics[width=9.0cm]{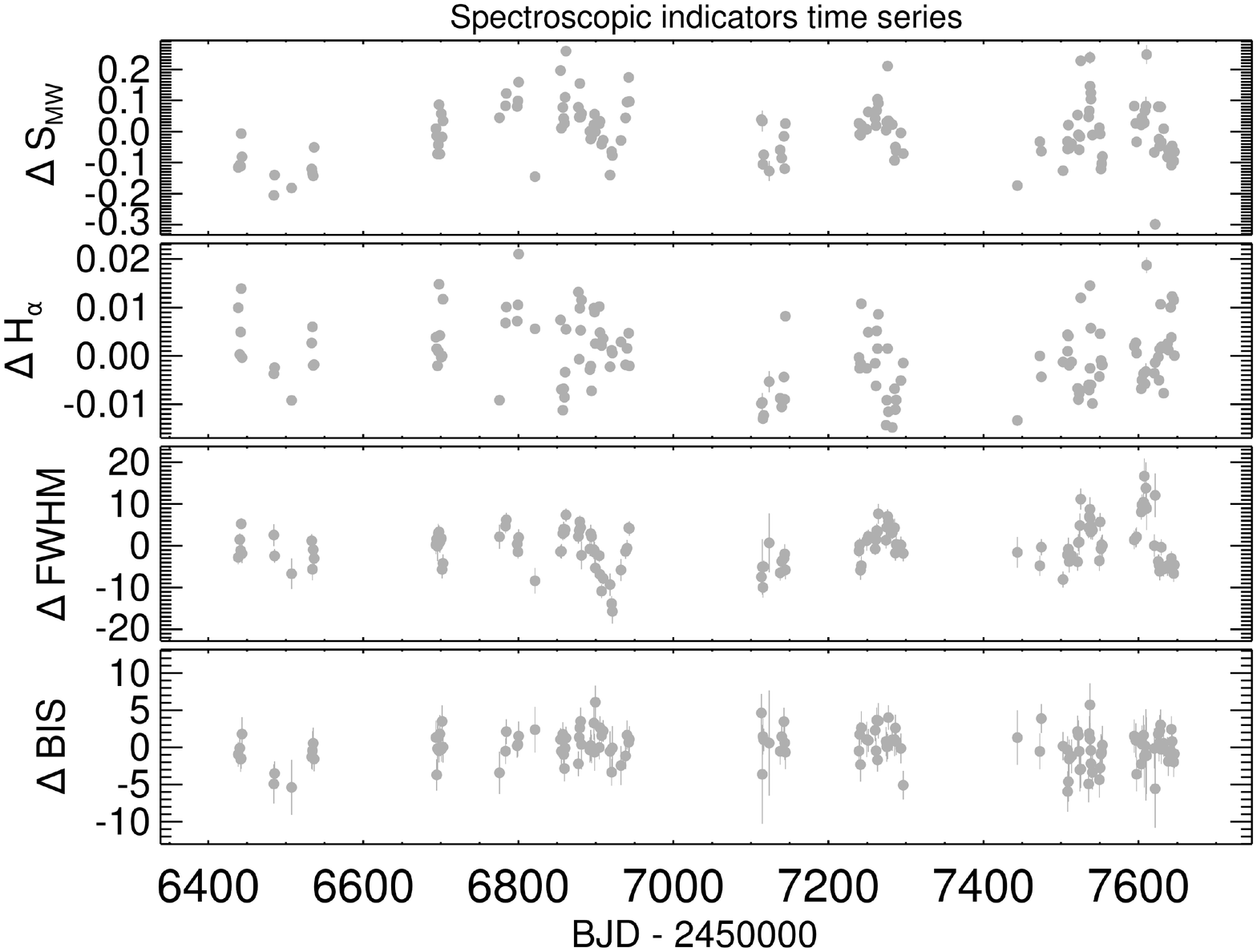}
	\caption{Time-series of the spectroscopic indicators. From top to bottom S$_{MW}$ index, H$_{\alpha}$ index, FWHM (in ms$^{-1}$) and bisector span (in ms$^{-1}$) time-series.}
	\label{SPEC_series}
\end{figure}

Figure~\ref{SPEC_series} shows the time-series for the four spectroscopic indicators used for the activity analysis: The  S$_{MW}$ index, H$_{\alpha}$ index, FWHM and bisector span time-series. For the four quantities we have 140 measurements distributed along 3.3 years. For the S$_{MW}$ index data we measure a mean value of 0.735 (corresponding to a $\log_{10}(R'_\textrm{HK})$ of --5.5) with an RMS of the data of 0.094 and a typical uncertainty of the individual measurement of 0.012. Using the activity relationships of \citet{Masca2015, Masca2016} we come to expect a rotation period around 80 $\pm$ 20 days.  For the H$_{\alpha}$ index we measure a mean value of 0.41607 with an RMS of the data of 0.00729 and a typical uncertainty of 0.00070. For the case of the FWHM time series we measure a mean FWHM of 3.83 km s$^{-1}$ with an RMS of the data of 5.32 m s$^{-1}$ and a typical uncertainty of 2.0 m s$^{-1}$. For the Bisector Span we measure a mean Span of -6.72 m s$^{-1}$ with an RMS of the data of 2.28 m s$^{-1}$ and a typical uncertainty of 2.0 m s$^{-1}$.

\begin{figure}
	\includegraphics[width=9.0cm]{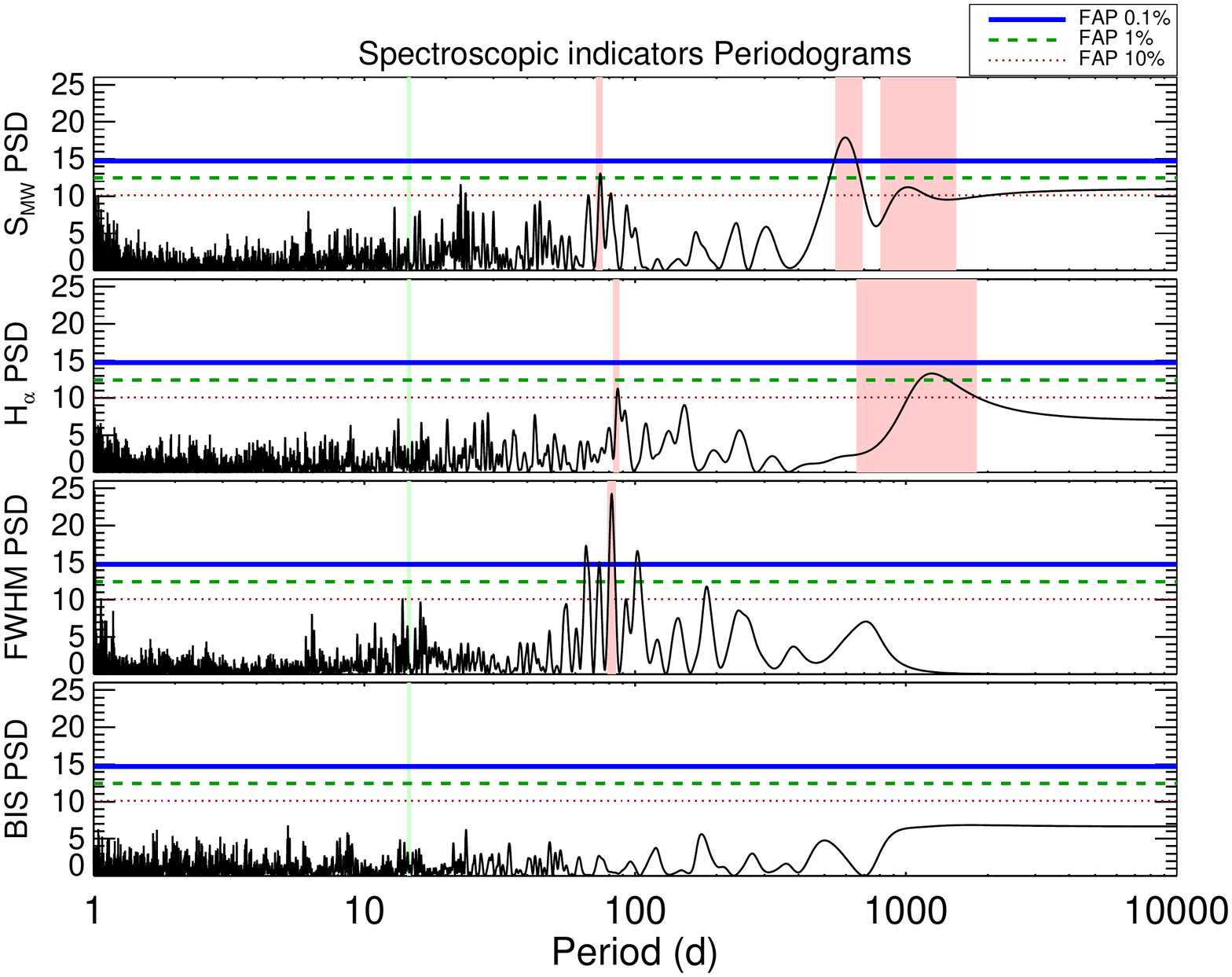}
	\caption{Periodograms of the spectroscopic indicators. From top to bottom S$_{MW}$ index, H$_{\alpha}$ index, FWHM and bisector span time-series. The light-red shaded regions show the detected periodicities. The light green shaded region shows the position of the 14.6 d signal detected in RV.}
	\label{SPEC_periodograms}
\end{figure}

\begin{figure*}
\begin{minipage}{0.33\textwidth}
        \centering
        \includegraphics[width=6.cm]{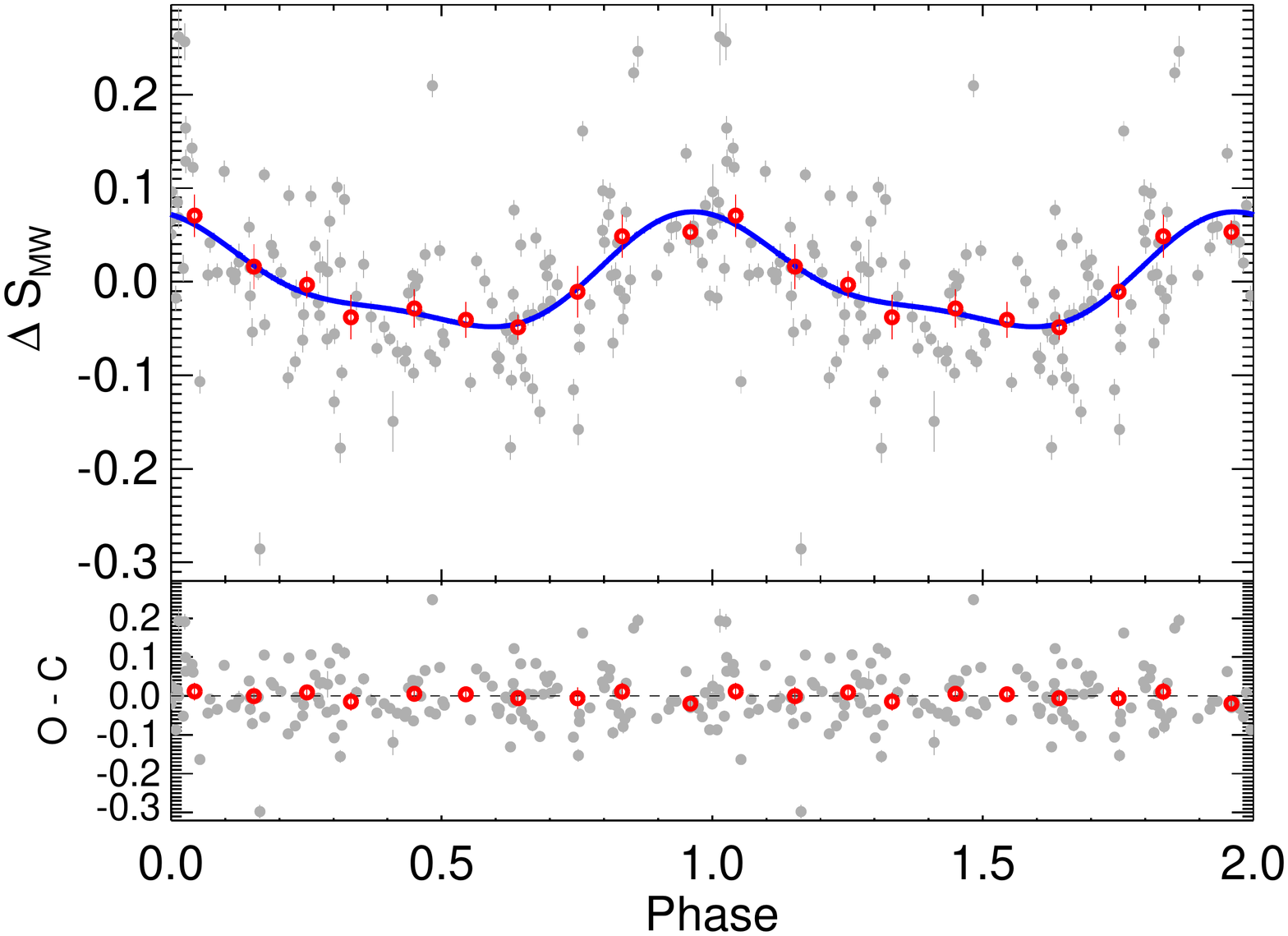}
\end{minipage}%
\begin{minipage}{0.33\textwidth}
        \centering
        \includegraphics[width=6.cm]{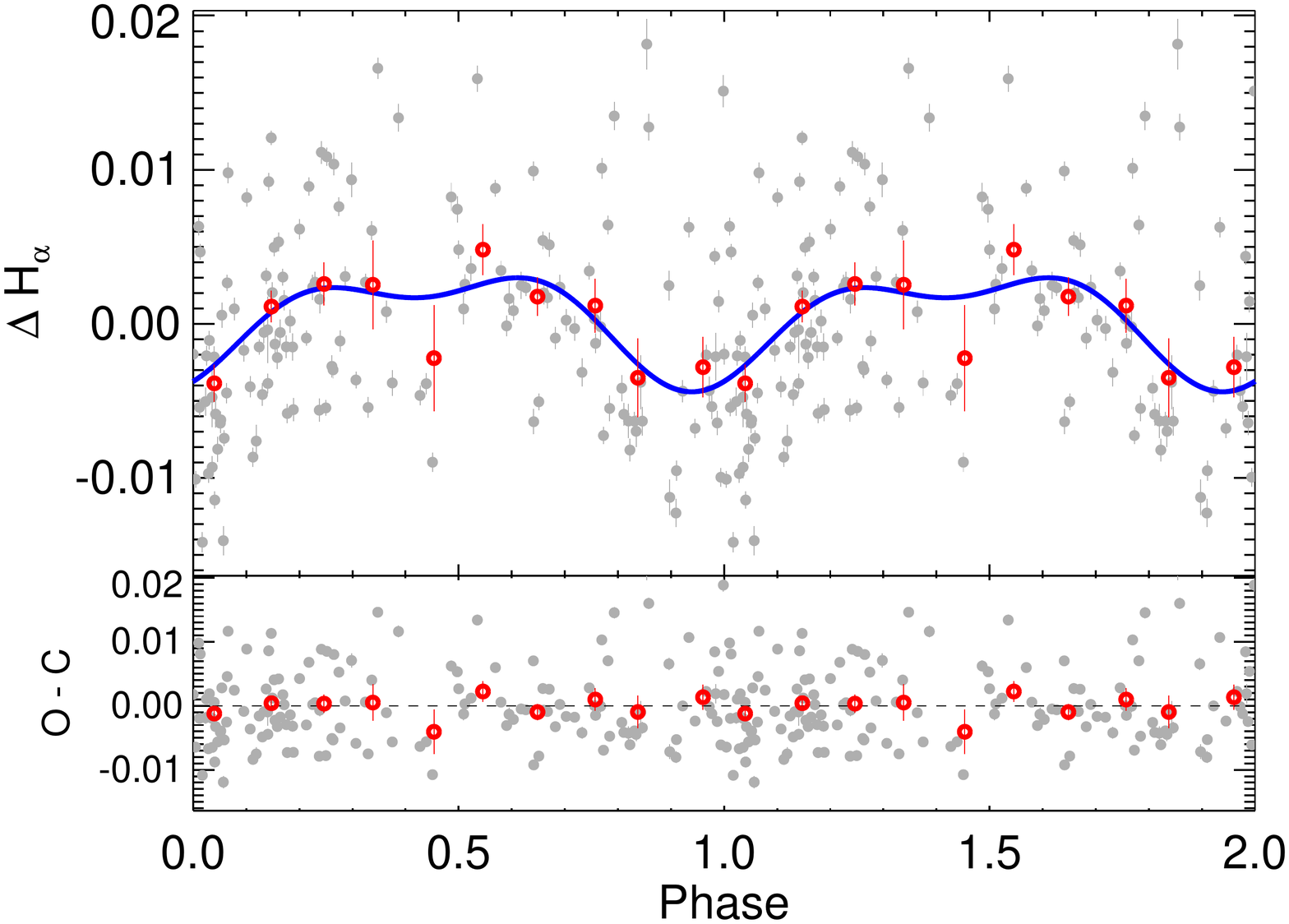}
\end{minipage}%
\begin{minipage}{0.33\textwidth}
        \centering
		\includegraphics[width=6.cm]{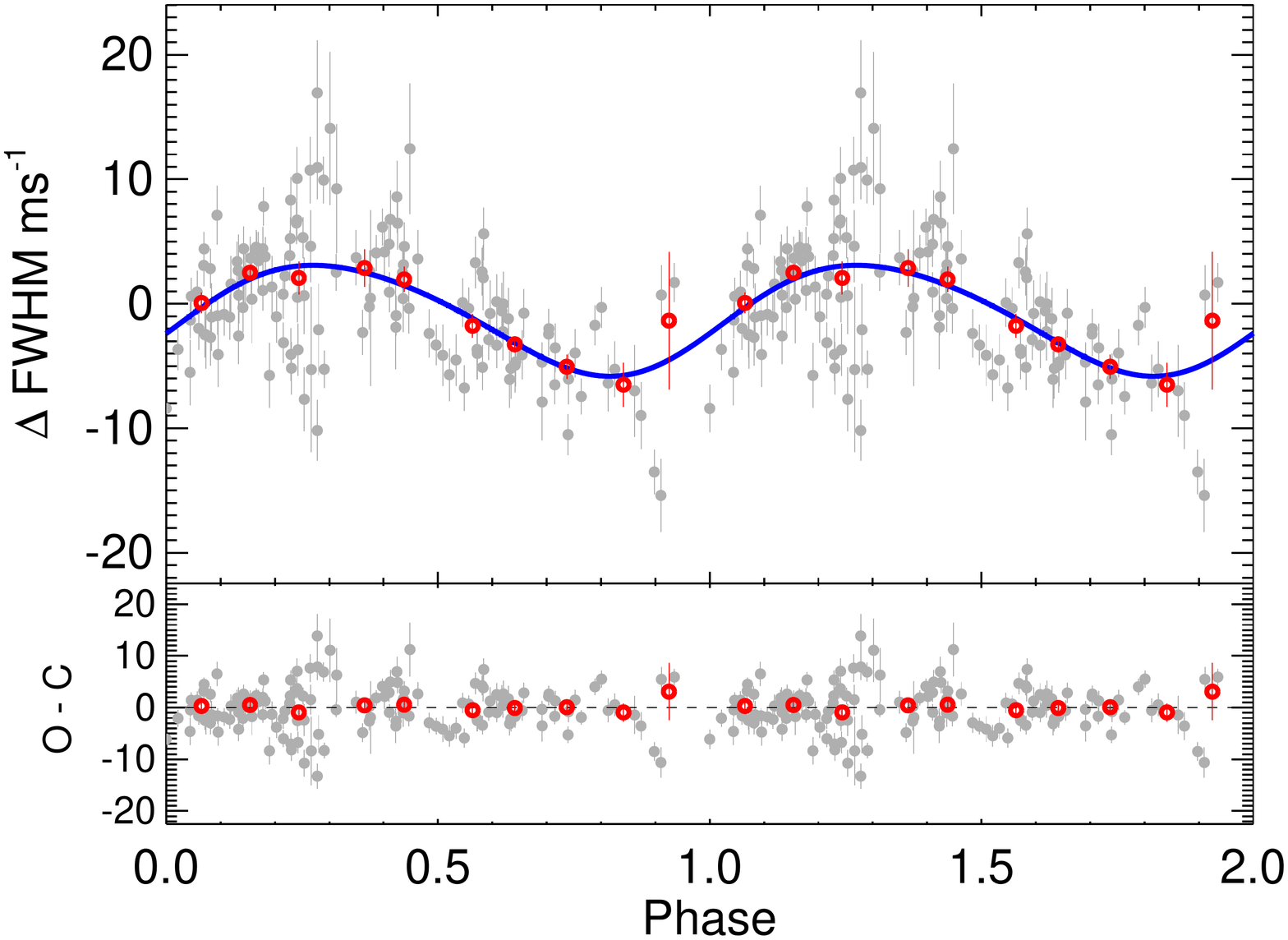}
\end{minipage}%

\caption{Phase folded curve for the S$_{MW}$ index  time series using the 74 d period, for the detrended H$_{\alpha}$ index time-series using the 85 d period, and for the FWHM time-series using the 82 d period. Grey dots are the individual measurements after subtracting the long term variations. Red dots are the same points binned in phase with a bin size of 0.1. The error bar of a given bin is estimated using the weighted standard deviation of binned measurements divided by the square root of the number of measurements included in this bin. The blue solid line shows the best fit to the data using a double harmonic sinusoidal fit.}
\label{SPEC_rot}
\end{figure*}

Figure~\ref{SPEC_periodograms} shows the periodograms for the four spectroscopic indicators. We see the sign of a long term signal in both the S$_{MW}$ and H$_{\alpha}$ indexes. In the S$_{MW}$ index periodogram it appears as a double peak at $\sim$ 1.7 yr and $\sim$ 3.2 yr, with large uncertainties in both cases. In the H$_{\alpha}$ index periodogram it appears a single peak at $\sim$ 3.4 yr. These signals would be compatible with short magnetic cycles, either flip-flop or global cycles \citep{BerdyuginaUsoskin2003, Moss2004,Masca2016}. The longer period peaks in both indexes seem to be the mark of the same $\sim$ 3.4 yr signal, while the short period peak in the S$_{MW}$ index might be an artefact caused by the signal being extremely asymmetric. It might also be a flip-flop cycle, combined with a 3.4 yr global cycle. More observations will be needed to confirm and constrain the cycle length and nature. Our 3.3 yr baseline of observations is too short to reach a unique conclusion on the nature of these long term signals. Apart from the long term signals, three shorter period signals appear in the S$_{MW}$, H$_{\alpha}$ and FWHM time series of 74, 85 and 82 days respectively. Those signals are consistent with the expected rotation period of the star. The S$_{MW}$ index signal is a 73.7 $\pm$ 2.2 days signal with a semi-amplitude of 0.0413 $\pm$ 0.0020 detected with a FAP < 1\%. The signal in the H$_{\alpha}$ index is a 85.2 $\pm$ 2.2 days signal with a semi-amplitude of 0.00217 $\pm$ 0.00014 detected with a FAP < 10\%. Finally the FWHM signal has a period of 82.0 $\pm$ 3.1 days and a semi-amplitude 4.03 $\pm$ 0.38 ms$^{-1}$, and a FAP < 0.1\%. Figure~\ref{SPEC_rot} shows the phase folded curves of the three signals.  

No more significant signals are found in any of the time-series after subtracting these signals.

\subsection*{Photometry}

\begin{figure}
	\includegraphics[width=9.0cm]{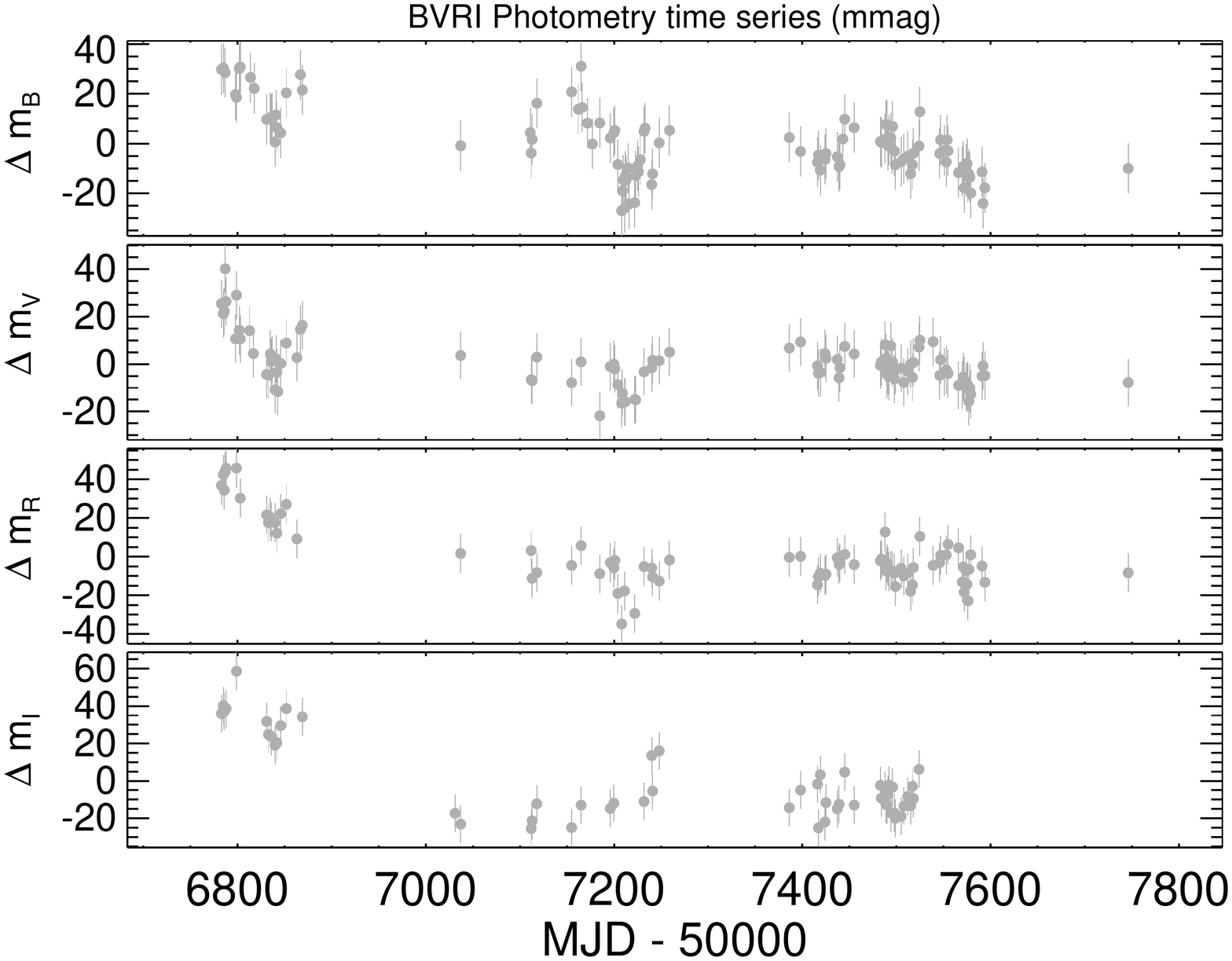}
	\caption{Time-series of the BVRI photometry. From top to bottom m$_{B}$, m$_{V}$, m$_{R}$ and m$_{I}$ time-series.}
	\label{BVRI_series}
\end{figure}

In order to confirm the possible activity-induced signals present in the RV time-series and in the time-series of spectroscopic indicators we analyse the BVRI photometric time-series. Figure~\ref{BVRI_series} and Table~\ref{phot_tab} show the available data. All series show the presence of some long-term evolution which is still not well covered by our observation baseline. 

\begin {table}
\begin{center}
\caption {Available photometric measurements \label{tab:phot_tab}}
    \begin{tabular}{ l | l  l l l l l l l l l l } \hline
Band  & N$_{Obs}$ & T$_{Span}$ (yr) & RMS (mmag) & $\sigma$ (mmag) \\ \hline
B & 108 & 2.6 & 13.6 & 10.1 \\
V & 98 & 2.6 & 10.4 & 10.0\\
R & 82 & 2.6 & 16.5 & 10.0\\
I & 55 & 2.0 & 21.1 & 10.0\\ \hline

\end{tabular}  
\label{phot_tab}
\end{center}

\end {table}

\begin{figure}
	\includegraphics[width=9.0cm]{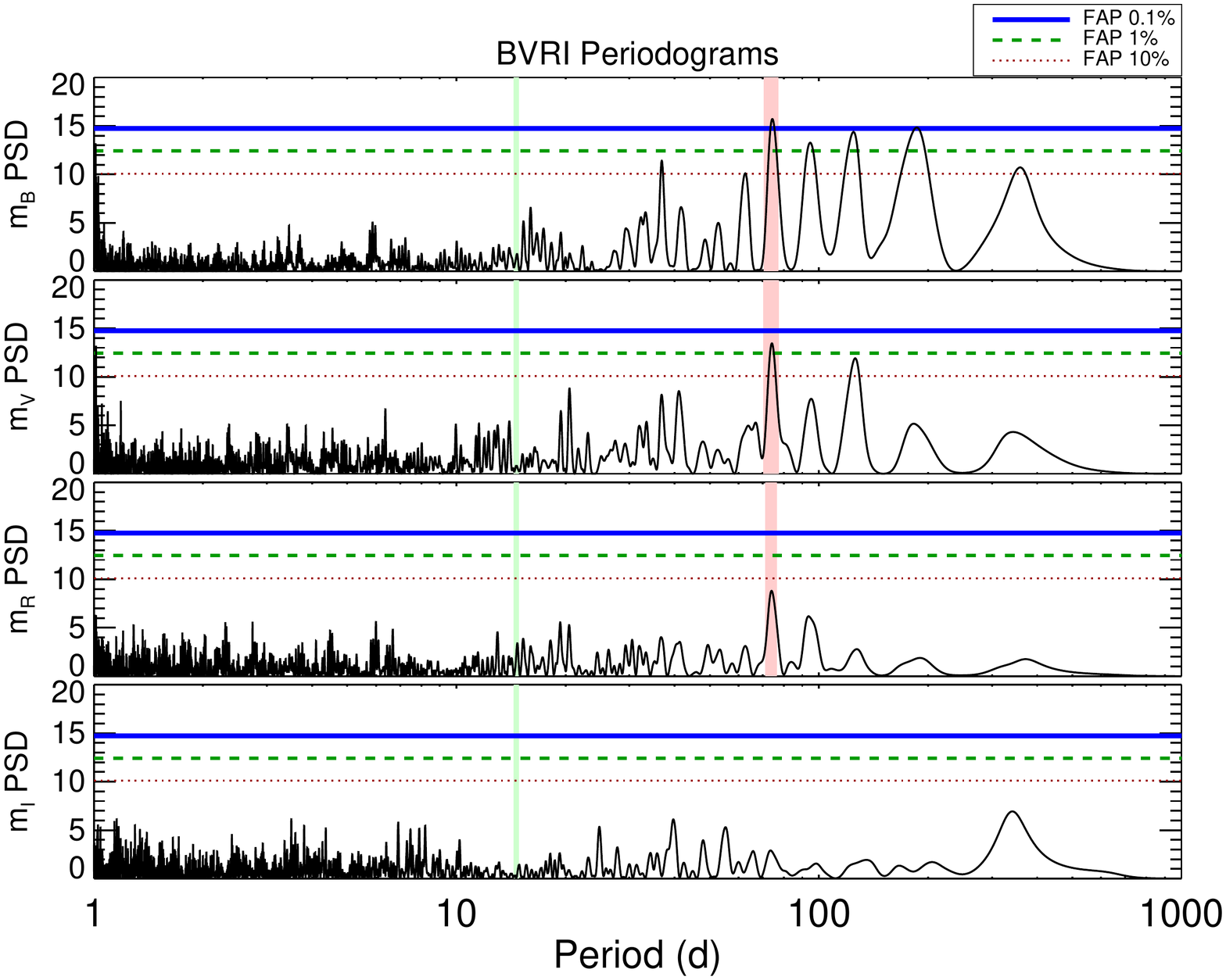}
	\caption{Periodograms for the BVRI photometry. From top to bottom m$_{B}$, m$_{V}$, m$_{R}$ and m$_{I}$ periodograms. The light-red shaded regions show the detected periodicities. The light green shaded region shows the position of the 14.6 d signal detected in RV.}
	\label{BVRI_periodograms}
\end{figure}

\begin{figure*}
\begin{minipage}{0.33\textwidth}
        \centering
        \includegraphics[width=6.cm]{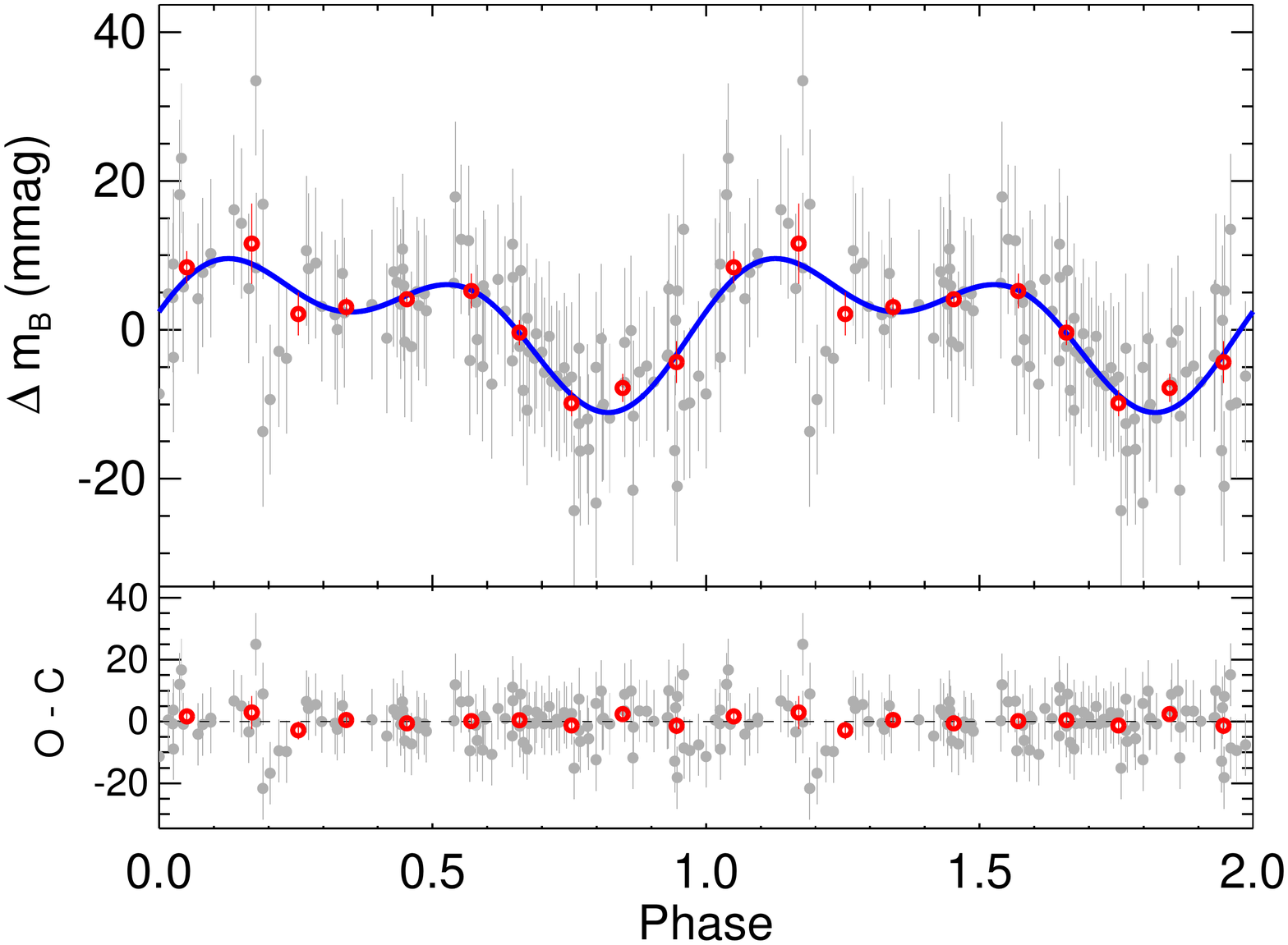}
\end{minipage}%
\begin{minipage}{0.33\textwidth}
        \centering
        \includegraphics[width=6.cm]{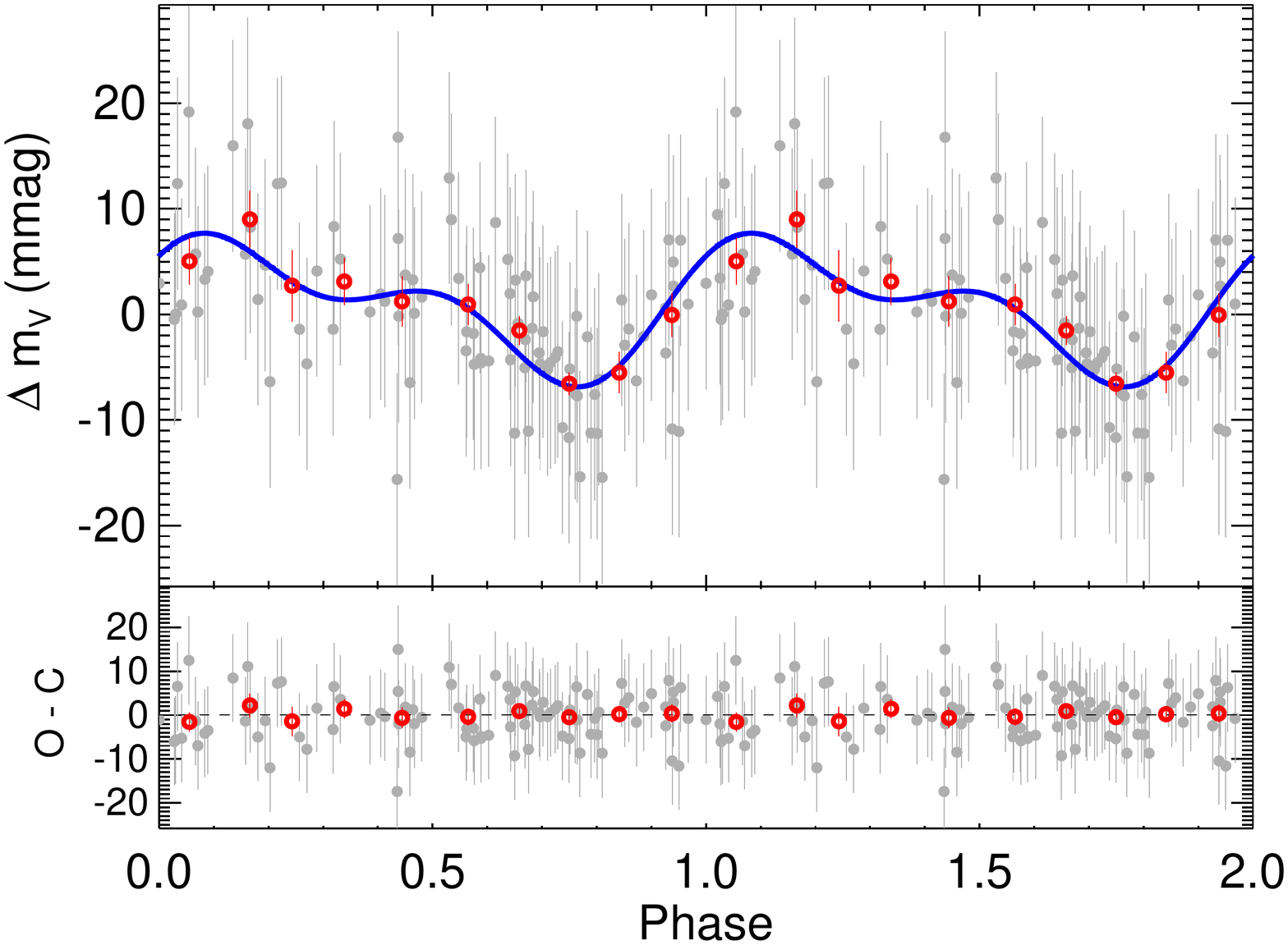}
\end{minipage}%
\begin{minipage}{0.33\textwidth}
        \centering
		\includegraphics[width=6.cm]{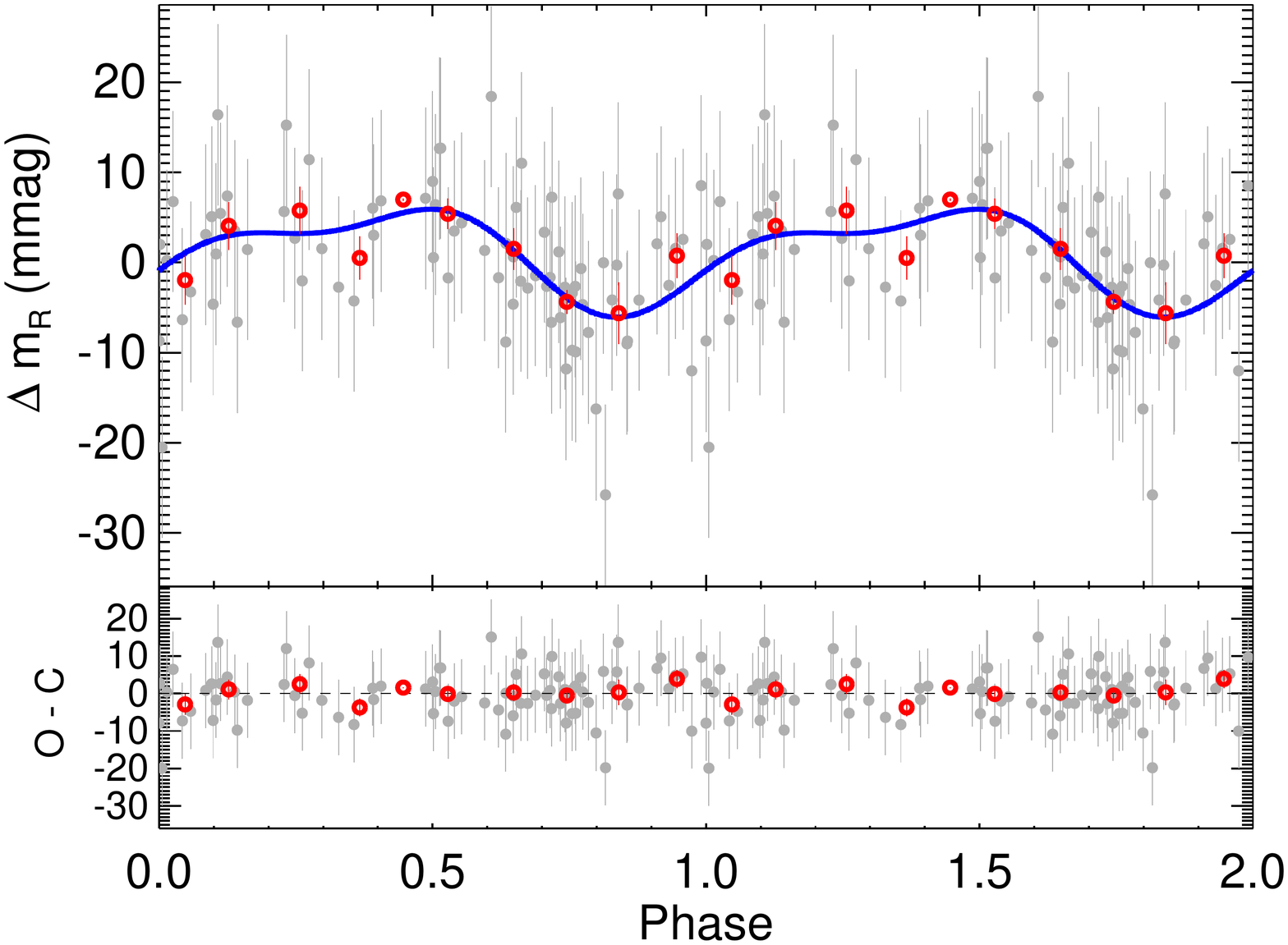}
\end{minipage}%

\caption{Phase folded curve of the data for the B, V and R filter time series using the 74 days periodicity. Grey dots are the individual measurements after subtracting the long term variations. Red dots are the same points binned in phase with a bin size of 0.1. The error bar of a given bin is estimated using the weighted standard deviation of binned measurements divided by the square root of the number of measurements included in this bin. The blue solid line shows the best fit to the data using a double harmonic sinusoidal fit.}
\label{BVR_phot_rot}
\end{figure*}

We subtract the long term evolution using a polynomial fit and analyse the residuals of the fit for periodic signals. Figure~\ref{BVRI_periodograms} shows the periodograms of the time series for the four filters. Several significant signals appear in the periodograms of the different time series, but we find one common periodicity for the m$_{B}$, m$_{V}$ and m$_{R}$ series at $\sim$ 74 d. We do not find any significant signal in the m$_{I}$ series periodogram. Being the I band the smallest dataset it is not surprising. We find a very significant (FAP < 0.1\%) signal at 73.9 $\pm$ 3.4 days with a semi-amplitude of 12.1 $\pm$ 0.3 mmag in the B band time series (Fig.~\ref{BVR_phot_rot}, left panel), a significant (FAP < 1\%) signal at 73.9 $\pm$ 3.7 days with a semi-amplitude of 8.2 $\pm$ 0.3 mmag in the V band time series (Fig.~\ref{BVR_phot_rot}, middle panel) and a less significant (FAP < 10\%) signal at 73.6 $\pm$ 2.8 days with a semi-amplitude of 7.8 $\pm$ 0.3 mmag for the R band time series (Fig.~\ref{BVR_phot_rot}, right panel). No more significant signals are found in the V and R band series after subtracting the detected signals. For the case of the B band there is a remaining signal at $\sim$ 1 yr with a second peak at $\sim$ 0.5 yr which seem to be an artefact.

\section{Interpretation of the detected signals} \label{int}

The previous analysis unveiled several signals at different time-scales. Some of them are common among different datasets. Table~\ref{signals} shows the measured signals for the different datasets of GJ 625. We can identify 6 different signals: At $\sim$3 yr, at $\sim$1.6 yr, at $\sim$85 d, at $\sim$82 d, at $\sim$74 d and at $\sim$14.6 d. 

\begin {table*}
\begin{center}
\caption {Detected signals in the different datasets of GJ 625. Parameters have been calculated using least squares minimization. \label{tab:signals}}
    \begin{tabular}{ l  l  l l l l l l l l l l l l } \hline
Dataset  & 14.6 d & Amp  &  74 d & Amp & 82 d & Amp  \\ \hline
RV CCF & 14.629 $\pm$ 0.069 d & 1.85 $\pm$ 0.13 ms$^{-1}$ & 74.7 $\pm$ 1.9 d & 1.64 $\pm$ 0.18 ms$^{-1}$ \\
RV TERRA & 14.629 $\pm$ 0.077 d & 1.65 $\pm$ 0.18 ms$^{-1}$ \\
S$_{MW}$ & & &  73.7 $\pm$ 2.2 d & 0.0413 $\pm$ 0.0020 \\
H$_{\alpha}$&\\
FWHM & & &&  & 82.0 $\pm$ 3.1 d & 4.03 $\pm$ 0.38 ms$^{-1}$\\
m$_{B}$ & &&  73.9 $\pm$ 3.4 d & 12.1 $\pm$ 0.3 mmag\\
m$_{V}$ & & & 73.9 $\pm$ 3.7 d & 8.2 $\pm$ 0.3 mmag\\
m$_{R}$ & &&  73.6 $\pm$ 2.8 d & 7.8 $\pm$ 0.3 mmag\\
 \hline \\ \\
 \hline
 Dataset  & 85 d & Amp & 1.7 yr & Amp & 3.4 yr & Amp \\ \hline
RV CCF &\\
RV TERRA & 85.9 $\pm$ 2.8 d & 1.58 $\pm$ 0.18 ms$^{-1}$\\
S$_{MW}$ &  & & 1.7 $\pm$ 0.2 yr &  0.08162 $\pm$ 0.0024 & 3.2 $\pm$ 1.0 yr & 0.05849 $\pm$ 0.0024\\
H$_{\alpha}$& 85.2 $\pm$ 2.2 d & 0.00217 $\pm$ 0.00014  & & &  3.4 $\pm$ 1.6 yr & 0.00593 $\pm$ 0.00017\\
FWHM &\\
m$_{B}$ & \\
m$_{V}$ & \\
m$_{R}$ & \\
 \hline
\label{signals}
\end{tabular}  
\end{center}
\end{table*}

The 3.4 yr signal appears as a significant peak in the H$_{\alpha}$ index time-series and as a less significant one in the S$_{MW}$ index time-series indicating a probable common origin. This signal is probably the fingerprint of the magnetic cycle of the star. A 3.4 yr  cycle is short for what one could expect in solar type stars, but is in the range of previous measurements in M-dwarfs \citep{Robertson2013b, Masca2016}. The signal at 1.7 yr could have different interpretations. On one hand, we could either be seeing the first harmonic of the 3 yr signal, because of the non-sinusoidal shape of the cycle \citep{Waldmeier1961,Baliunas1995}. On the other hand, it could be a flip-flop cycle, i.e. a cycle of spatial rearrangement of active regions \citep{BerdyuginaUsoskin2003, Moss2004}. The short baseline of observations proves to be a problem to give a correct interpretation of these signals. We could be measuring the global cycle and a flip flop cycle, or the global cycle (3 yr) and its first harmonic or even a flip-flop cycle and its first harmonic. At this stage, with our baseline being as close to the measured cycle length we cannot rule out the possibility of having an artefact of a poorly constrained long term signal. Many years of data are probably needed before giving a definitive answer. 

The group of signals at 74, 82 and 85 days that appear in the different spectroscopic indicators, photometric series and in the radial velocity series, are probably related to the stellar rotation. Given the activity level of the star ($\log_{10}(R'_\textrm{HK})$ = --5.5 $\pm$ 0.2 ) we expect the rotation period to be around 80 $\pm$ 20 days \citep{Masca2015, Masca2016}, and the semi amplitude of the induced radial velocity signal to be $\sim$ 1.5 ms$^{-1}$ \citep{Masca2017b}. \citet{Scandariato2017} found a similar periodicity using a pooled variance analysis in the Ca II H\&K and H$_{\alpha}$ fluxes, and the V-band photometric time series. The 5 signals around 74 days (RV, CCF, S$_{MW}$, m$_{B}$, m$_{V}$ and m$_{R}$) give us a strong indication of the stellar rotation, probably in a region close to the equator. The decline in amplitude when moving to redder bands in photometry might be an indication that we are measuring a modulation based on photospheric inhomogeneities, for which the contrast gets reduced when going to redder wavelength. The signal at 85 days seen in the H$_{\alpha}$ index and RV TERRA time series probably shows the rotation period at higher latitudes, giving us a hint on the differential rotation of this star. Being a signal only present in the H$_{\alpha}$ and RV variations we are probably seeing a modulation based only on chromospheric inhomogeneities. The 82 days signal seen in the FWHM could be both a signal based on inhomogeneities in an intermediate latitude, or an average of all the measured inhomogeneities along all latitudes. The two different RV induced signals for the two algorithms could imply that the two different RV algorithms are sensitive to the effect of different groups of inhomogeneities. While the CCF measurements seems to go in line with the photometric and Ca II  H\&K variations, the TERRA measurements appear coherent the H$_{\alpha}$ variations. 

The signal at 14.6 days appears in both analyses of the RV time-series with parameters that are consistent with each other. There is no evidence of signals at this period in any of the available activity proxies (see Table~\ref{signals}).

As a second test we measured the Spearman correlation coefficient between the S$_{MW}$, the H$_{\alpha}$ index, the FWHM, the bisector span and the radial velocities. We do not find a strong correlation between any of these quantities and the raw radial velocities. The only significant one being between the FWHM and the RVs, specially with the CCF ones. All the correlation coefficients remain non significant if we subtract 74 and 85 days signals, isolating the 14.6 days signal (see Table.~\ref{tab:rv_corre}). For the case of the FWHM vs RV correlation we see the coefficient going down significantly. When we subtract the 14.6 d signal isolating the 74 d or 85 d signals  we see the RV vs S$_{MW}$ correlation coefficients going up and the RV vs FWHM correlation coefficient recovering the raw value. This constitutes a new evidence of the stellar origin of the 74 d and 85 d signals, and of the planetary origin of the 14.6 d one. Although the correlation coefficients are not very high in any case, it seems clear that isolating the 14.6 d signal reduces the correlation between the different activity proxies and the RV measurements, while isolating the 74 d or 85 d signal does not, and even increases it for the S$_{MW}$.

\begin {table*}
\begin{center}
\caption {Activity - Radial-velocity correlations. The parenthesis value indicates the significance of the correlation. Note: Long term variations of activity indicators have been subtracted. 
 \label{tab:rv_corre}}
    \begin{tabular}{ l | c  c c c } 
    \hline
Parameter & Raw data & 14.6 d signal &  74 d signal & 85 d signal\\ \hline
S$_{MW}$ vs V$_{R} CCF$ & 0.08 (1$\sigma$) & -0.05 ($\textless$1$\sigma$)  & 0.14 (2$\sigma$)& \\
S$_{MW}$ vs V$_{R} TERRA$ & 0.09 (1$\sigma$) & -0.01 ($\textless$1$\sigma$)  & & 0.13 (2$\sigma$)\\
H$_{\alpha}$ vs V$_{R} CCF$ & -0.08 (1$\sigma$) & -0.08 (1$\sigma$) & 0.02 ($\textless$1$\sigma$) \\
H$_{\alpha}$ vs V$_{R} TERRA$ & 0.01 ($\textless$1$\sigma$)  & -0.03 ($\textless$1$\sigma$)  & &  0.05 ($\textless$1$\sigma$) \\
FWHM vs V$_{R} CCF$ & 0.23 (2$\sigma$) & 0.14 (1$\sigma$) & 0.22 (2$\sigma$)\\
FWHM vs V$_{R} TERRA$ & 0.15 (2$\sigma$) & 0.06 ($\textless$ 1$\sigma$) & & 0.14 (2$\sigma$)\\
BIS vs V$_{R} CCF$ & 0.04($\textless$ 1$\sigma$) & 0.05 ($\textless$ 1$\sigma$)& 0.06 ($\textless$ 1$\sigma$)\\
BIS vs V$_{R} TERRA$ & -0.01 ($\textless$1$\sigma$) & 0.01 ($\textless$1$\sigma$) & & -0.01 ($\textless$1$\sigma$) \\ 
\hline
\end{tabular}  
\end{center}
\end {table*}

\begin{figure}
	\includegraphics[width=9cm]{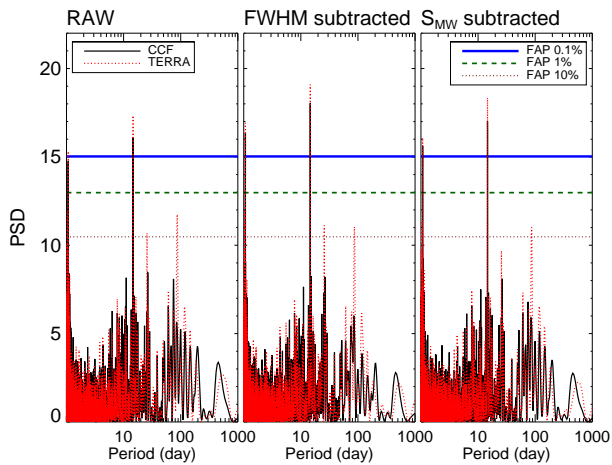}
	\caption{Periodograms for the radial velocity after removing the correlation with the different activity diagnostic tools that showed a significant correlation coefficient. From left to right there is the periodogram for the original data, the periodogram after detrending against the FWHM, and against the S$_{MW}$ index.}
	\label{rv_nocorr}
\end{figure}

Following this idea, we subtract the linear correlation between the radial velocity and the two activity diagnostic indexes that showed a significant correlation. When doing this we see, that the strength of the 14.6 d signal remains constant, or even gets increased, while the significance of the 74 d and 85 d signals gets slightly reduced in all cases (see Fig.~\ref{rv_nocorr}).

\begin{figure}
	\includegraphics[width=9cm]{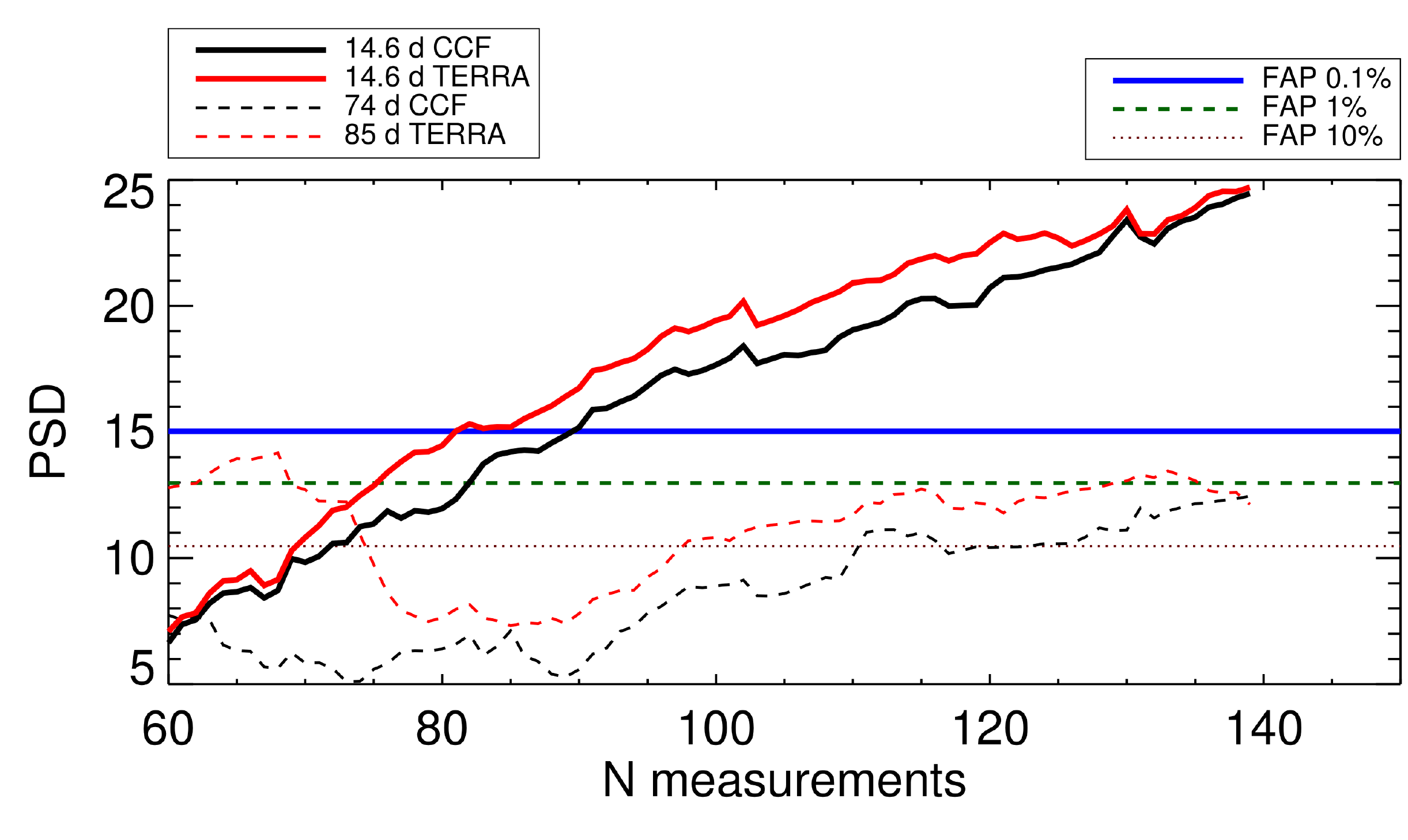}
	\caption{Evolution of the significance of the detections for the isolated signals. Thick lines show the 14.6 days signal, dashed lines represent the 74 d and 85 d signals. Black lines represent the CCF signals, red lines show the TERRA signals.}
	\label{evol}
\end{figure}

Keplerian signals are deterministic and consistent in time. When measuring one signal, it is expected to find the significance of the detection increasing steadily with the number of observations, as well as the measured period being stable over time. However, in the case of activity related signal this is not necessarily the case. As the stellar surface is not static, and the configuration of active regions may change with time, changes in the phase of the modulation and in the detected period are expected \citep{Affer2016, Masca2017, Mortier2017}. Even the disappearance of the signal at certain seasons is possible.  To study the evolution of both signals we perform a simultaneous fit of the detected periodicities in each of the time series, and then use the derived parameters to subtract the contribution of one of them leaving the other "isolated". We then perform the stacked periodograms using a very narrow frequency window around each of the signals. Fig.~\ref{evol} shows the evolution of the PSD of the detection of both isolated signals. Once we gain enough signal to noise, the 14.6 d signal increases steadily with the number of measurements. On the other hand the behaviour of the 74 d and 85 d signals is more erratic, especially for the case of the 85 days signal in the TERRA data. This is consistent with the attributed stellar origin, although we cannot rule out that the fluctuations are created by the lack of a sufficient signal to noise ratio for those signals. 

Of the different significant radial-velocity signals detected in our data, it seems clear that the one at 14.6 d has a planetary origin, while the signals at 74 d and 85 d have stellar activity  origin. 

\begin{figure}
	\includegraphics[width=9cm]{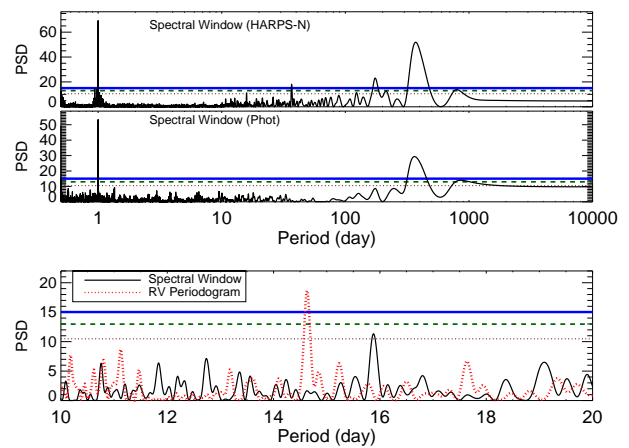}
	\caption{Spectral windows for the HARPS data (top panel) and the photometric data (mid panel). Bottom panel shows the RV periodogram (red dotted line) compared to the spectral window (black solid line) in the range of periods around the signal for the planet candidate.}
	\label{wind}
\end{figure}

Finally, an analysis of the spectral window ruled out that the peak in the periodogram attributed to the keplerian signal is an artefact of the time sampling. No features appear at 14.6 d days even after masking the oversaturated regions of the power spectrum. The region around 70-90 days is more complicated in the HARPS-N spectral window, casting some doubts on those periodicities, but that region is perfectly clean in the spectral window for the photometric time series. That suggests that the signals that we attributed to rotation in the RVs and spectroscopic indicators are real, but it also might explain the differences between the various indicators and the two different RV time series trough spectral leakage of the real signal~\citep{Scargle1982}.  We also recognize a prominent feature at 36.8 days, which should be treated with caution when searching for additional planets in the future. Figure~\ref{wind} shows the spectral windows of the HARPS-N and photometric time series, along with a comparison between the RV periodogram around the 14.6 d signal and the spectral window in the same region. Following \citet{Rajpaul2016}, we tried to re-create the 14.6 days signal by injecting the P$_{\rm Rot}$ signal along with a second signal at P$_{\rm Rot}$/2 at 1000 randomized phase shifts with a white noise model. Additionally we added a second set of signals at 3 yr and 1.5 yr to account for the possible effect of the magnetic cycle. We were never able to generate a signal at 14.6 days, or any significant signal at periods close to 14.6 days. It seems very unlikely that any of the signals are artefacts of the sampling. The process on the other hand created many spurious peaks at periods between 35 and 120 days. Future observations should take into account the possibility of having aliases and artefacts of the rotation arising in a wide range of periodicities.

\section{MCMC modelling of the GJ 625 b planet}

The analysis of the radial-velocity time series and of the activity indicators leads us to conclude that the best explanation of the observed data is the existence of a planet orbiting the star GJ 625 at the period of 14.6 d. The best solution comes from a super-Earth with a minimum mass of 2.8 M$_{\oplus}$ orbiting at 0.078 AU of its star. 

In order to quantify the uncertainties of the orbital parameters of the planet, we perform a bayesian analysis using the code {\sc ExoFit}~\citep{Balan2009} as outlined in \citet{Masca2017}. This code follows the Bayesian method described in ~\citet{Gregory2005,Ford2005,FordGregory2007}. A single planet can be modelled using the following formula:
\begin{equation}
	v_i = \gamma - K [\sin( \theta(t_i+\chi P) + \omega ) + e \sin \omega ]
\end{equation}

where $\gamma$ is system radial velocity; $K$ is the velocity semi-amplitude equal to $2\pi P^{-1} (1-e^2)^{-1/2} a \sin i$; $P$ is the orbital period; $a$ is the semi-major axis of the orbit; $e$ is the orbital eccentricity; $i$ is the inclination of the orbit;
$\omega$ is the longitude of periastron; $\chi$ is the fraction of an orbit, prior to the start of data taking, at which periastron occurs (thus, $\chi P$ equals the number of days prior to $t_i=0$ that the star was at periastron, for an orbital period 
of $P$ days); and $\theta(t_i + \chi P)$ is the angle of the star in its orbit relative to periastron at time $t_i$, also called the true anomaly.

To fit the previous equation to the data we need to specify the six model parameters, $P$, $K$, $\gamma$, $e$, $\omega$ and $\chi$.
Observed radial-velocity data, $d_i$, can be modelled by the equation: $d_i = v_i + \epsilon_i + \delta$ \citep{Gregory2005}, where $v_i$ is the calculated radial velocity of the star and $\epsilon_i$ is the uncertainty component arising from accountable but unequal measurement error which are assumed to be normally distributed. The term $\delta$ contains any unknown measurement error. Any noise component that cannot be modelled is described by the term $\delta$. The probability distribution of $\delta$ is chosen to be a Gaussian distribution with finite variance $s^2$. Therefore, the combination of uncertainties $\epsilon_i + \delta$ has a Gaussian distribution with a variance equal to $\sigma_i^2 + s^2$~\citep[see][for more details]{Balan2009}.

\begin{figure*}
	\includegraphics[width=18cm]{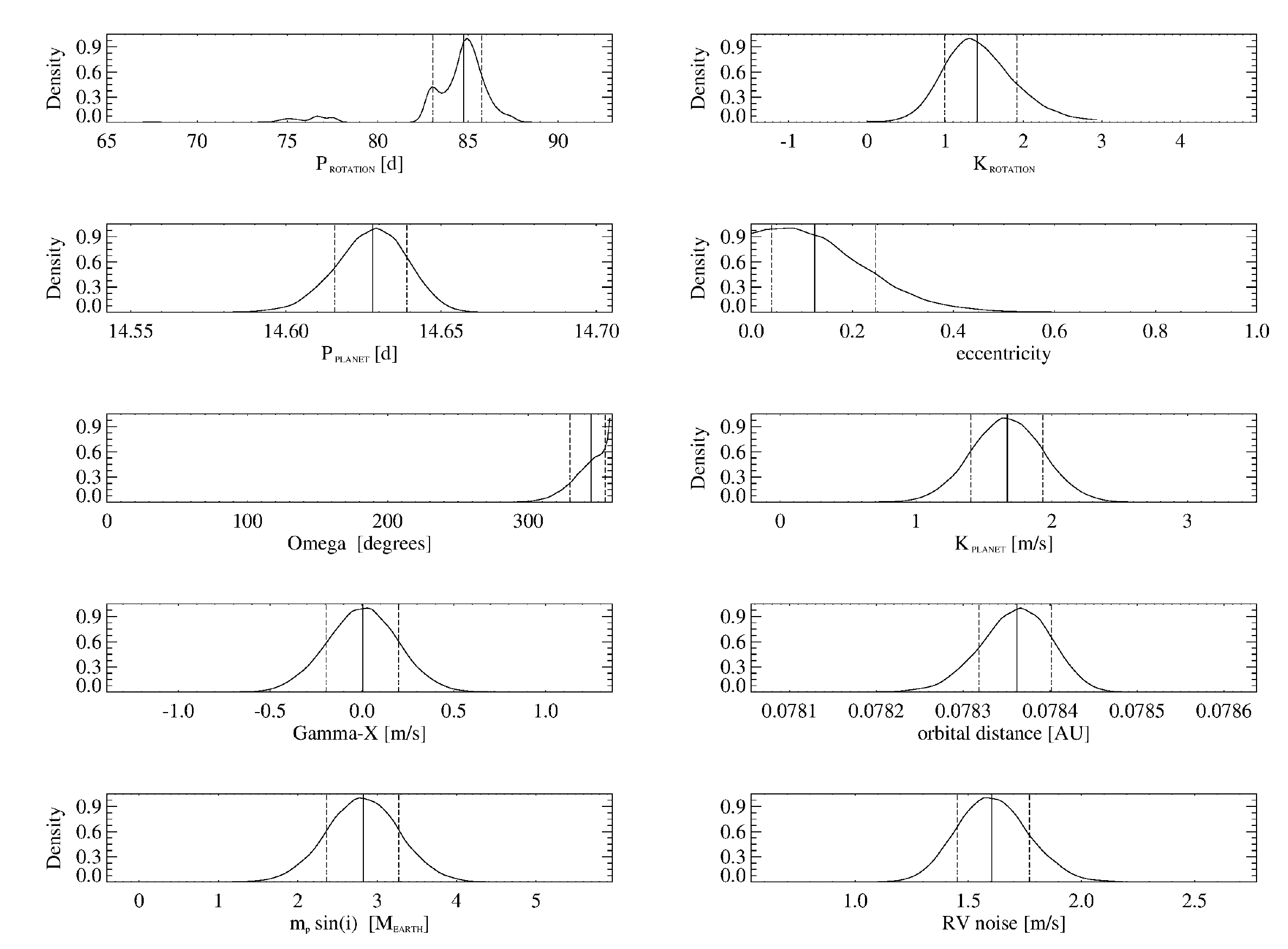}
\caption{Posterior distribution of model parameters of the planet companion of the M dwarf star GJ~625 using the RV measurements given by the TERRA pipeline. Vertical dashed line show the median value of the distribution and the dotted lines the 1-$\sigma$ values.}
\label{mcmc}
\end{figure*}

\begin{figure*}
	\includegraphics[width=18cm]{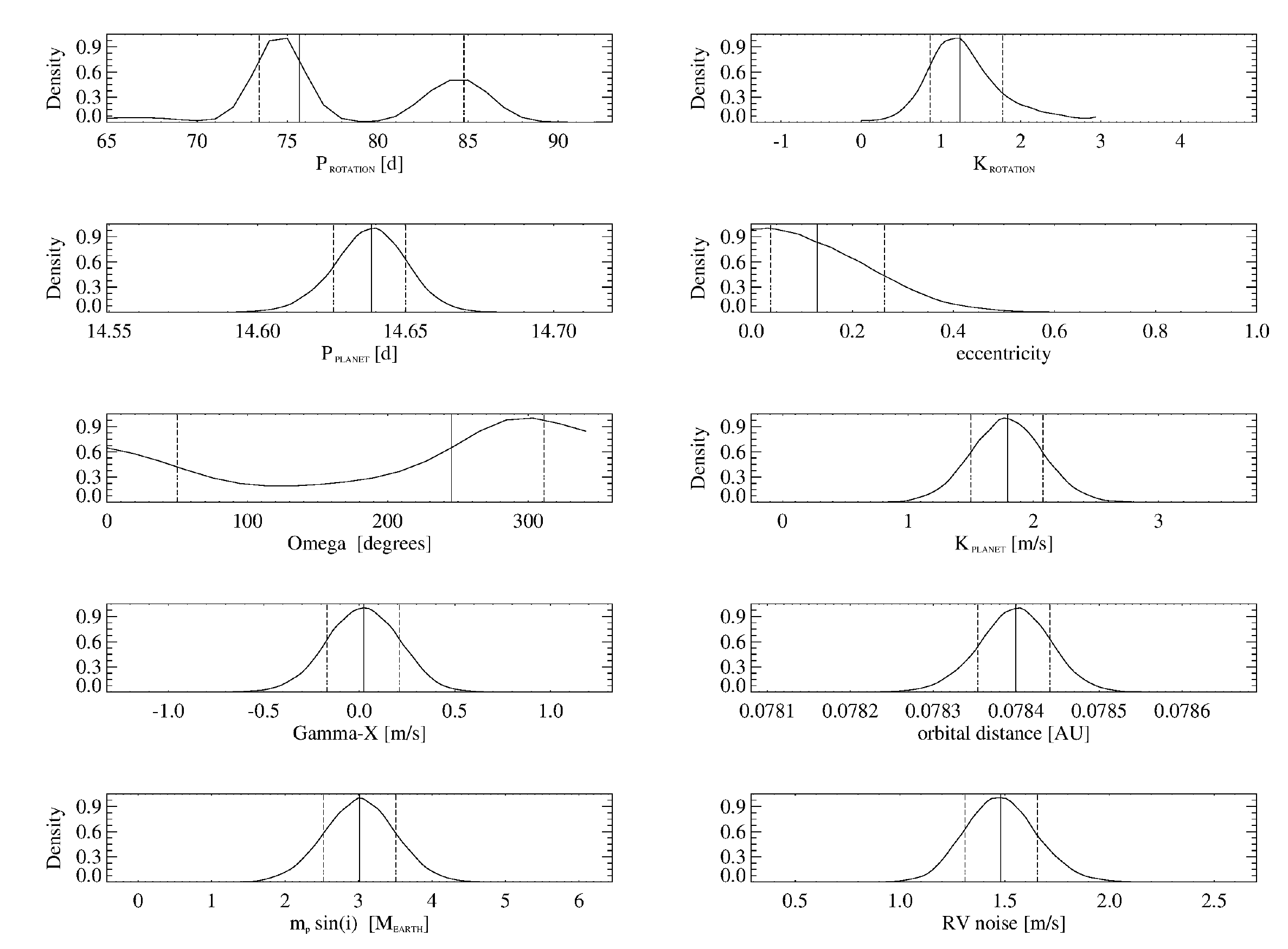}
\caption{Posterior distribution of model parameters of the planet companion of the M dwarf star GJ~625 using the RV measurements given by the CCFs. Vertical dashed line show the median value of the distribution and the dotted lines the 1-$\sigma$ values.}
\label{mcmc_ccf}
\end{figure*}

\begin{figure*}
\begin{minipage}{0.5\textwidth}
        \centering
        \includegraphics[width=9.cm]{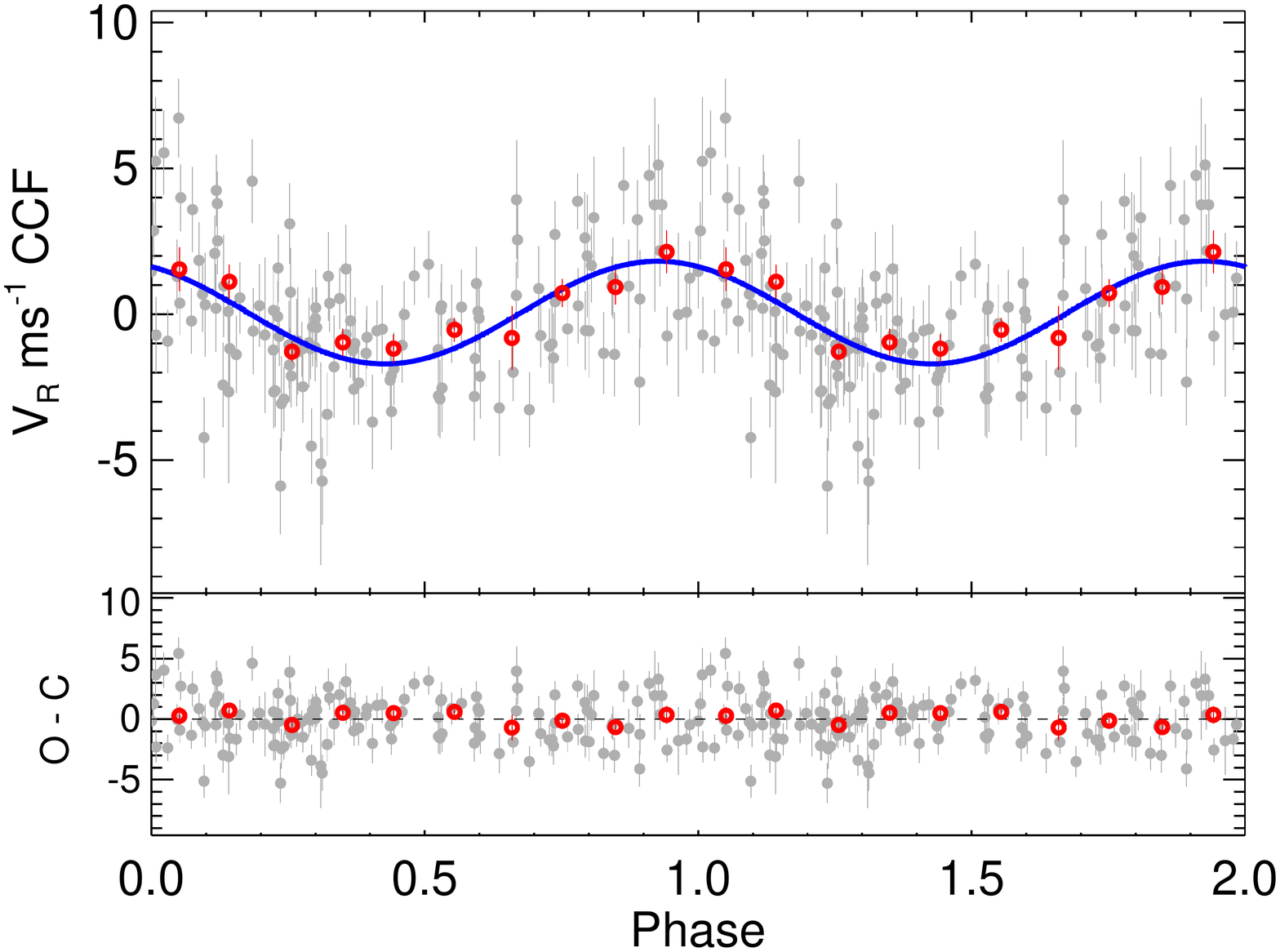}
\end{minipage}%
\begin{minipage}{0.5\textwidth}
        \centering
        \includegraphics[width=9.cm]{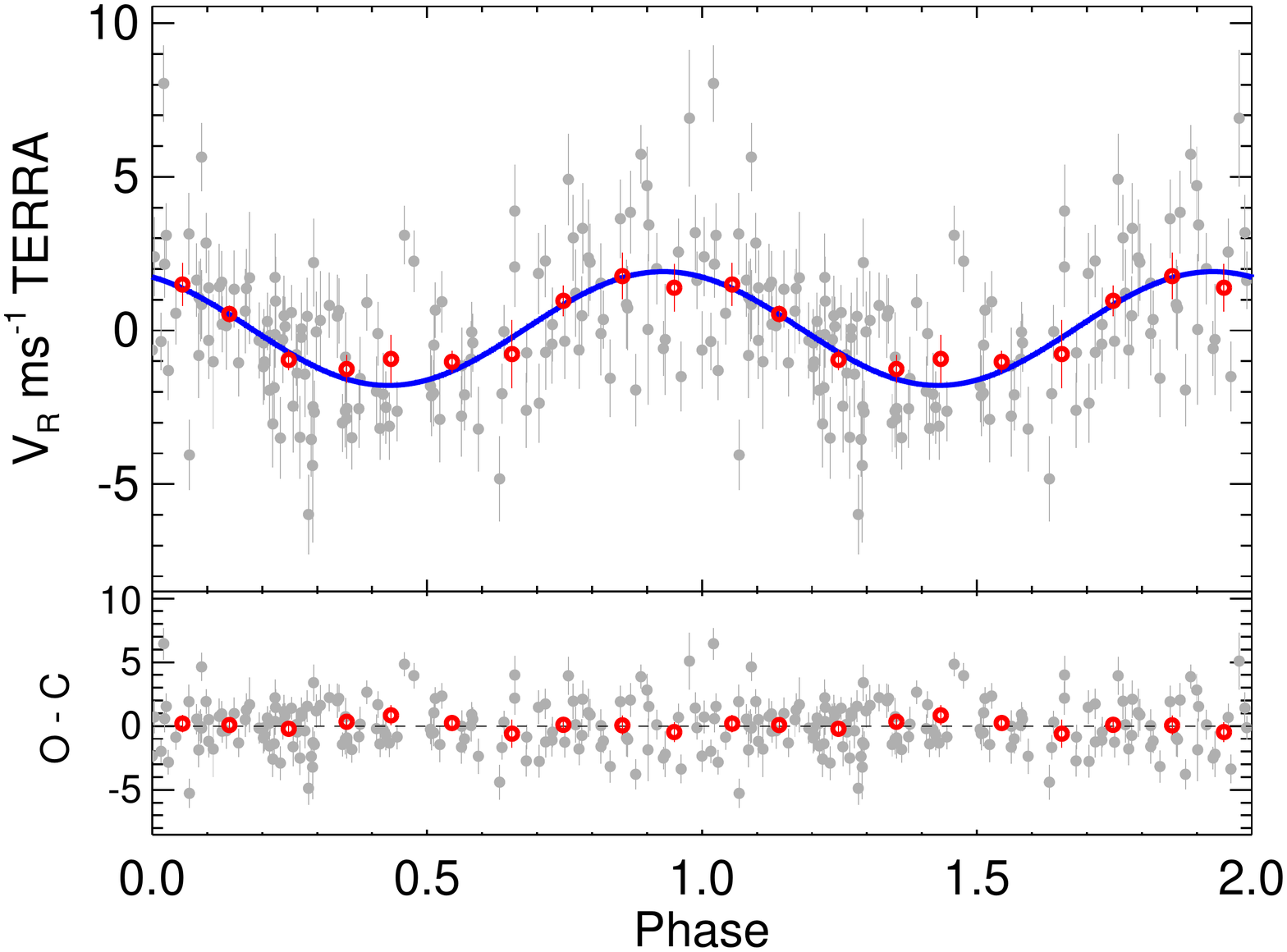}
\end{minipage}%
\caption{Phase folded curve of the planetary signal detected in GJ 625 using the parameters of the MCMC model. Left panel shows the CCF measurements, right panel the TERRA measurements. Grey dots show the measurements after subtracting the detected activity induced signals. Red dots are the same points binned in phase with a bin size of 0.1. The error bar of a given bin is estimated using the weighted standard deviation of binned measurements divided by the square root of the number of measurements included in this bin. Blue line shows the best fit to the data using a Keplerian model.}
\label{RV_pla_def}
\end{figure*}

In Table~\ref{mcmc_par} we show the final parameters and uncertainties obtained with the MCMC bayesian analysis with the code {\sc ExoFit}. We performed a a simultaneous fit of the planetary signal and the activity induced signal using both the CCF and the TERRA data. The obtained parameters are compatible in both datasets, except for $\omega$ and $\chi$. However in those cases the uncertainties are large, suggesting that more data might be needed to better define the solution. Fig.~\ref{mcmc} and ~\ref{mcmc_ccf} show the probability densities for all the parameters in both situations and Fig.~\ref{RV_pla_def} shows the best fit to the data obtained for the RV signal attributed to the planet candidate GJ 625 b. The TERRA data gives a smaller RV amplitude and smaller mass and a smaller RMS of the residuals. The noise factor in both cases exceeds the typical uncertainties of our data, indicating the presence of unaccounted signals in the radial velocity data. The smaller RMS of the residuals after the fit of the TERRA data -- despite showing a higher RV noise factor -- and the tighter parameter results lead us to favour the model given by the TERRA data. The RMS of the remaining residuals is 1.8 m s$^{-1}$, smaller than the HADES noise contribution estimated in \citet{Perger2017}.

\begin{table*}
\begin{center}
\caption {MCMC parameters and uncertainties\label{mcmc_par}}
\begin{tabular}{l | c c c | c c c | c }

 \multicolumn{1}{r}{} & \multicolumn{3}{c}{\textit{TERRA}} & \multicolumn{3}{c}{\textit{CCF}}\\
\hline
Parameter & Value & Upper error & Lower error & Value & Upper error & Lower error & Prior\\
\hline
$P_{\rm planet}$ [d]             & 14.628 & $+$0.012 & $-$0.013 & 14.638 & $+$0.012 & $-$ 0.013 & 10.5 - 18.5 \\
$\gamma$ [ms$^{-1}+12850$]       & -0.11  & $+$0.18   & $-$0.18   & -0.07 & $+$ 0.19 & $-$ 0.20 & $-$3.0 - $+$3.0 \\
$e$                              & 0.13  & $+$0.12   & $-$0.09  & 0.13 & $+$ 0.13 & $-$ 0.09 & 0.0 - 0.99 \\
$\omega$ [deg]                   & 343.1  & $+$9.6   & $-$14.8  & 240 & $+$ 70 & $-$ 180 & 0.0 - 360.0 \\
$\chi$                           & 0.94+   & $+$0.026   & $-$0.043  & 0.26 & $+$ 0.45 & $-$ 0.18 & 0.0 - 0.99 \\
$K_{\rm planet}$ [ms$^{-1}$]     & 1.67   & $+$0.29   & $-$0.29  & 1.79 & $+$ 0.29 & $-$ 0.30 & 0.0 - 3.0 \\
$a$ [AU]                         & 0.078361 & $+$0.000044 & $-$0.000046  &  0.078399 & $+$ 0.000042& $-$ 0.000045 &-- \\
$m_p \sin i$ [M$_{\rm Earth}$]   & 2.82   & $+$0.51   & $-$0.51  & 3.02 & $+$0.49 & $-$ 0.50 & -- \\
 & & & & & & &   \\
$P_{\rm Rot}$ [d]             & 84.7 & $+$       1.3 & $-$        1.8 & 76.1 & $+$        8.9& $-$         3.0 &  65.0 - 95.0 \\
$K_{\rm Rot}$ [ms$^{-1}$]     & 1.41& $+$       0.53& $-$      0.47 &  1.25 & $+$      0.52& $-$       0.39 & 0.0 - 3.0 \\
 & & & & & & &   \\
RV noise [ms$^{-1}$]             & 1.61   & $+$0.18   & $-$0.18  & 1.48 & $+$ 0.18 & $-$ 0.17 & 0.0 - 3.0 \\
\hline
\end{tabular}  
\end{center}
\end{table*}

\section{Discussion}

We detect the presence of a planet with a semi-amplitude of 1.6 m s$^{-1}$ that, given the stellar mass of 0.3 M$_{\odot}$, converts to $m_{p} \sin i$ of 2.75 M$_{\oplus}$, orbiting with a period of 14.6 d around GJ 625, an M2-type star with a mean rotation period of around 78 d and an additional activity signal compatible with an  activity cycle of around 3 yr. We have seen hints of differential rotation, with the difference between the shortest and the longest period going up to 11 days.

\begin{figure}
	\includegraphics[width=9cm]{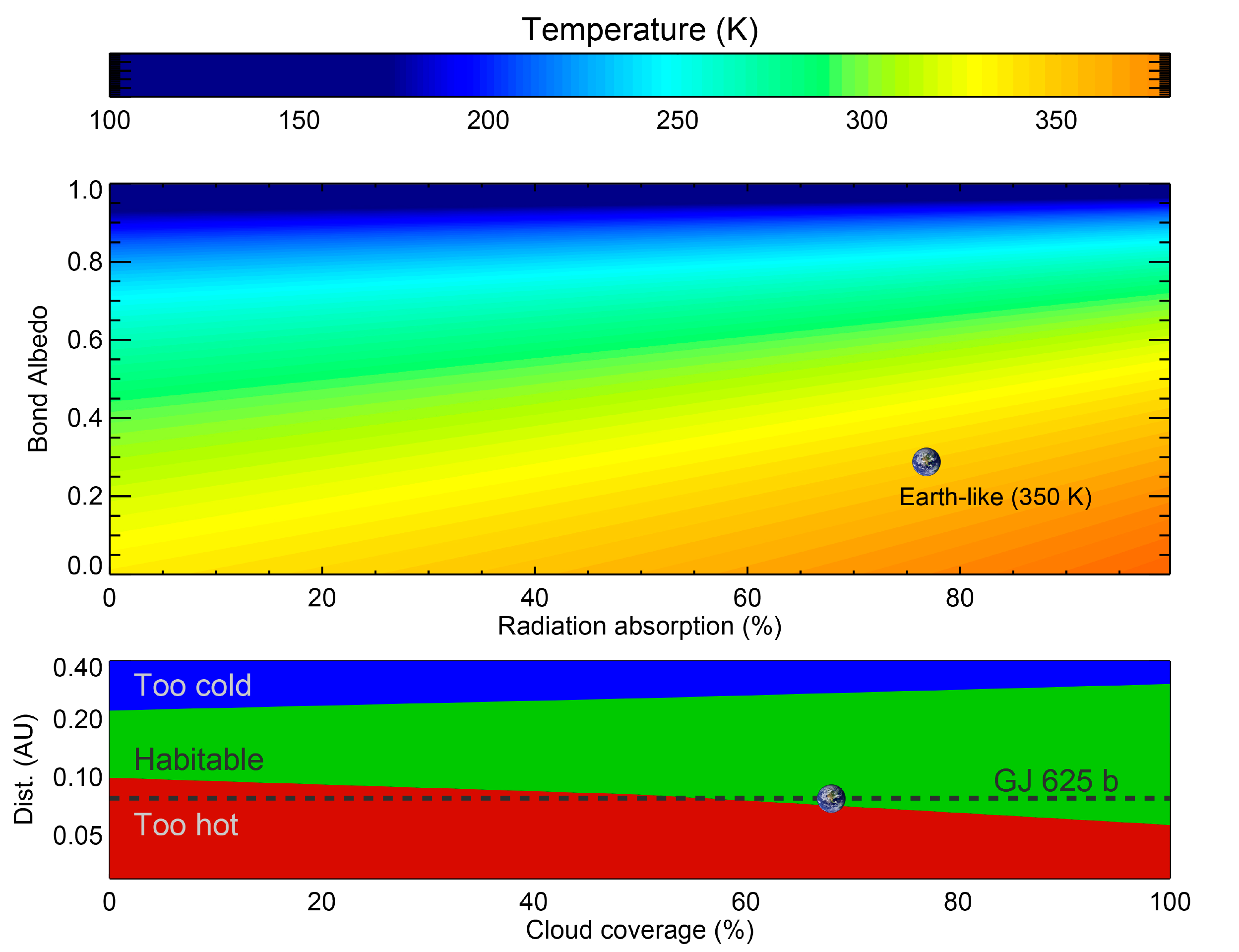}
\caption{Mean surface temperature of GJ 625 b as a function of albedo and the fraction of radiation absorbed by the atmosphere (top panel) and change of the habitable zone as a function of cloud coverage of the planet (bottom panel). The Earth symbol shows where it would be with an Earth-like albedo and greenhouse effect (top panel), and with an Earth-like cloud coverage (bottom panel).}
\label{temp}
\end{figure}

The planet is a small super-Earth at the edge of the habitable zone of its star. Using a basic estimation of the equilibrium temperature, and a correction using the greenhouse calculated to estimate the surface temperature, we estimate a mean surface temperature of 350 K for a Bond albedo A = 0.3 and Earth-like greenhouse effect. Surface temperatures close to earth surface temperature can be reached for many combinations of albedo and greenhouse effect. Following \citet{Kasting1993} and \citet{Selsis2007}, we perform a simple estimation of the habitable zone (HZ) of this star. The HZ would go from 0.099 to 0.222 AU in the narrowest case (cloud free model), and 0.057 to 0.305 AU in the broader one (fully clouded model). Figure ~\ref{temp} shows the distribution of surface mean temperatures for the different combination of bond albedo and green house levels and the evolution of the habitable zone following \citet{Selsis2007}. The estimation of the habitable zone performed by \citet{Kopparapu2013} for a cloud-free model would leave the most optimistic inner limit of the HZ at 0.088 AU. If we assume that the HZ evolves with the cloud coverage of the atmosphere of the planet in the same way as in \citet{Selsis2007}, then the most optimistic limit of the inner HZ would move down to 0.043 AU for a completely covered atmosphere, following almost exactly the same pattern as in ~\citet{Selsis2007}. We find that GJ 625 b might potentially host liquid water depending on its atmospheric conditions.

The habitability of planets in close orbits around low-mass stars is nowadays a subject of debate. Planets are probably tidally locked, and exposed to the strong magnetic field, flares and UV and X-ray irradiation of their parent stars. But even if those are strong arguments against their habitability, none of them is a definitive one. Depending on the planet composition and its own magnetic field it could be able to prevent all but a small atmospheric loss \citep{Vidotto2013, Zuluaga2013, AngladaEscude2016, Ribas2016, Bolmont2017}.

\begin{figure*}
\begin{center}
	\includegraphics[width=18cm]{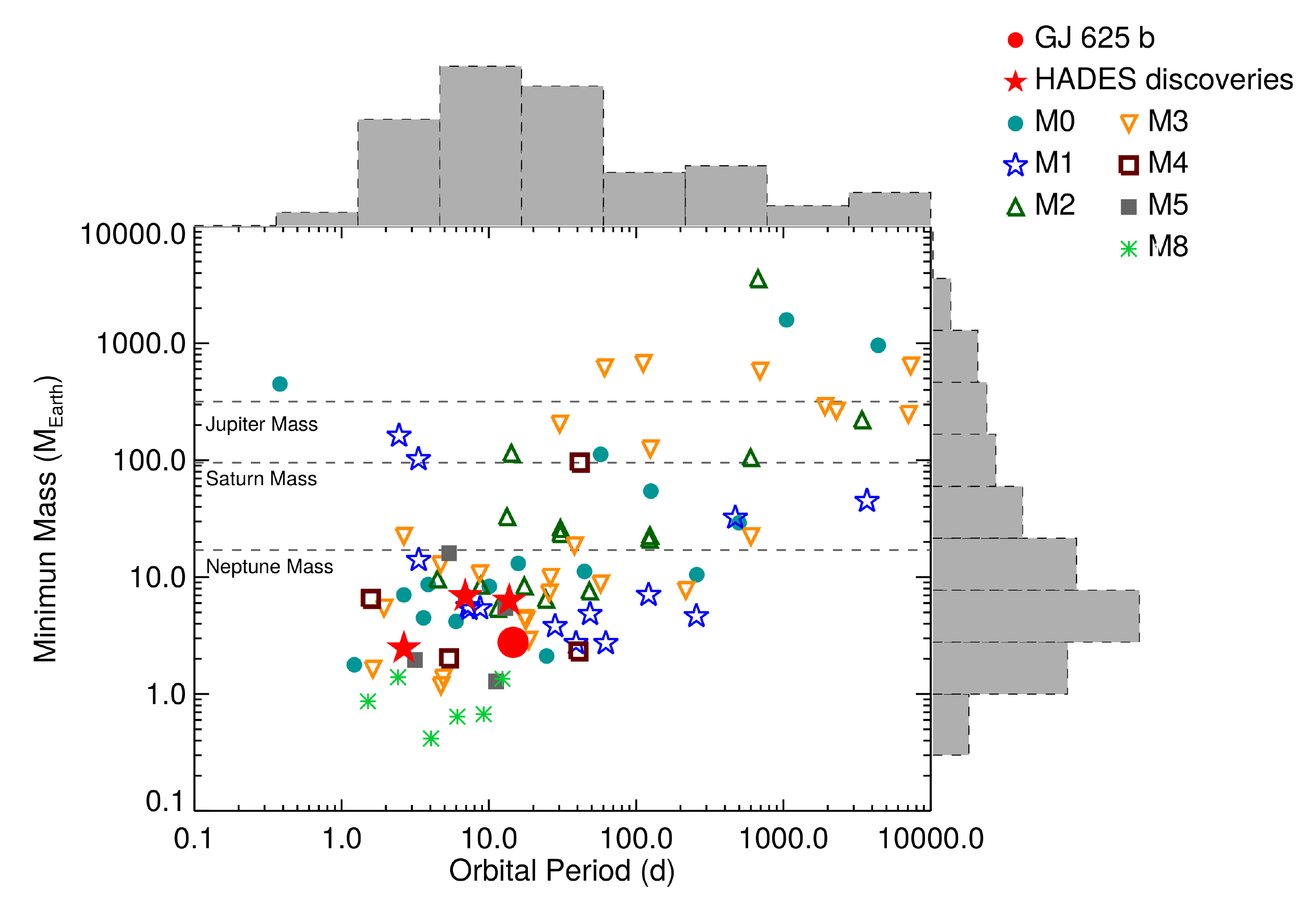}
	\caption{Minimum mass versus orbital period for the known planets with determined masses around M-dwarf stars  (using data from exoplanet.eu) divided by the spectral type of the host star. The red filled dot shows the position of GJ 625 b, and the red filled stars show the position of GJ 3998 b, c and GJ 3942 b \citep{Affer2016, Perger2017b}.  Horizontal dashed lines show the mass of the solar system planets for comparison. On the edges of the figure the distribution for each parameter is shown.}
	\label{period_mass}
	\end{center}
\end{figure*}

GJ 625 b is  in the lower part of the Mass vs Period diagram of  known planets around M-dwarf stars with measured dynamic masses (Fig.~\ref{period_mass}). The proposed projected mass would make it the lightest planet found around an M2 star to date. Of all the known planets around M-dwarfs with dynamical masses determined $\sim$ 52 \% are super-Earths or Earth-like planets ($\textless$ 10 M$_{\oplus}$) at periods shorter than 100 days. Fig.~\ref{period_mass} points out once more the lack of massive planets in close orbits around M-dwarfs, only $\sim$ 9 \% of detected planets at periods shorter than 10 days being heavier than 10 M$_{\oplus}$, with only one going over 1 M$_{Jup}$, as expected by the core-accretion formation model \citep{Laughlin2004}. There is also a lack of small planets ($\textless$ 3 M$_{\oplus}$) in close orbits ($\textless$ 10 days) around early M-dwarfs (M0-M2), with GJ 3998 being the only one. At periods longer than 10 days on the other hand there is a great abundance of this kind of planets.

The presence of one or more extra planets around the star GJ 625 cannot yet be ruled out at longer orbital periods and small RV amplitudes ( < 1.5 ms$^{-1}$ for periods shorter than 3 years), but finding them will not be easy. The rotation period, combined with the signs of differential rotation, would make difficult the detection of low amplitude planetary signals in the range of periods from 35 to 120 days, as stated at the end of Section~\ref{int}. Also considering the detected magnetic cycle it would be expected to find a RV signal at 1.5 yr or 3 yr eventually arising. The most promising candidates to continue searching would be very small mass planets at very close orbits, and small mass planets in the habitable zone of the star, at periods shorter than 30 days. 

The activity induced RV variations measured both using the CCF fitting and the TERRA pipeline are not consistent with each other. With the CCF fitting favouring a signal at 74 days, very similar to the ones measured in the B-,V- and R-band photometry and in the Ca II H\&K variations, while the TERRA pipeline favours a 85 d signal closest to the signal measured in the H$_{\alpha}$ index. Stellar induced signals are usually more complicated than planetary signals, and the long period and low amplitude of the signals ($\sim$ 1.6 ms$^{-1}$), combined with their non-sinusoidal shape, could easily lead to an incorrect modelling. The spectral window of the observation series might also be playing a part here. The region around 70-100 days is a complicated one, so it cannot be ruled out that it is affecting the 82 and 85 days measurements. If we assume that this is not the case, it could be that the change in the source of RV information (a selection of lines in the CCF vs the full spectrum in TERRA) is making the two algorithms sensitive to different active regions or even different depths of the stellar atmosphere. If this were the case the use of both algorithms for the same star could prove a very useful tool for the diagnosis of activity induced signals, maybe even for the case of earlier spectral types (G and K-type stars).   Further investigation using data from different stars would be needed.

GJ 625 has an estimated mass of 0.3 M$_{\odot}$, placing it close to the theoretical limit at which a star becomes fully convective. If the proposed 3 yr magnetic cycle could be confirmed it would place GJ 625 in the small group of very small mass stars which exhibit activity cycles despite theoretical predictions suggesting they should not ~\citep{Chabrier2006, Robertson2013, Masca2016,Wargelin2017}. The growing number of detected cycles in stars that are expected to be fully-convective, and therefore to not have a tachocline, supports the idea that the stellar dynamo might not be confined to the tachocline but instead distributed across the convection zone~\citep{Wright2016}. 

\section{Conclusions}

We have analysed 151 high resolution spectra along with photometric observations in 4 different bands to study the presence of planetary companions around the M-dwarf star GJ 625 and its stellar activity. For the study of this system, we extracted the radial velocity information in two different ways, on one hand using the cross correlation of the spectra with a numerical template, on the other hand by template matching the individual spectra with a high S/N template. 

We detected one significant radial-velocity signals, at a period of 14.6 d. A second one arises at 74-85 days, with different periods for the different algorithms. From the available photometric and spectroscopic information  we conclude that the 14.6 d signal is caused by a planet with a minimum mass of 2.82 M$_{\oplus}$ in an orbit with a semi-major axis of 0.078 AU. The planet is on the inner edge of the habitable zone, with its mean surface temperature very dependent on the atmospheric parameters. The short period of the planet makes it a potential transiting candidate. Following \citet{Winn2010} we find the transit probability of GJ 625 b in the range of 1.5 - 2.5 \%. Detecting the transits would give a new constraining point to the mass-radius diagram and provide further opportunity for atmospheric characterization. 

The second radial-velocity signal of period 74-85 d and semi amplitude of 1.6 ms$^{-1}$ is a magnetic activity induced signal related to the rotation of the star.  Using the photometric light curves and the time-series of spectral indicators we conclude that the rotation period of the star is $\sim$ 74 days, with evidence of differential rotation up to $\sim$ 85 days. We have seen that the measured rotation period and the amplitude of the rotation induced signal match the expected results of \citet{Masca2015, Masca2016, Masca2017b}. We also found evidence for a  magnetic cycle of $\sim$3 yr which would need future observations to be better constrained.

\section*{Acknowledgements}
We thank Guillem Anglada-Escudé for distributing the latest version of the TERRA pipeline. This research has made extensive use of the SIMBAD database, operated at CDS, Strasbourg, France and NASA's Astrophysics Data System. This work has been financed by the Spanish Ministry project MINECO AYA2014-56359-P. J.I.G.H. acknowledges financial support from the Spanish MINECO under the 2013 Ram\'on y Cajal program MINECO RYC-2013-14875. A.S.M acknowledges financial support from the  Swiss National Science Foundation (SNSF). M. P., I. R., J.C.M., E.H., A. R., and M.L. acknowledge support from the Spanish Ministry of Economy and Competitiveness (MINECO) and the Fondo Europeo de Desarrollo Regional (FEDER) through grant ESP2016-80435-C2-1-R. mGAPS acknowledges support from INAF trough the "Progetti Premiali” funding scheme of the Italian Ministry of Education, University, and Research and through the PREMIALE WOW 2013 research project. L. A. and G. M. acknowledge support from the Ariel ASI-INAF agreement N. 2015-038-R.0. The research leading to these results has received funding from the European Union Seventh Framework Programme (FP7/ 2007-2013) under Grant Agreement No. 313014 (ETAEARTH). G.S.\ acknowledges financial support from \lq\lq\ Accordo ASI--INAF\rq\rq\ n. 2013-016-R.0 July 9, 2013.

%
%
\bibliography{RHK_ref}

\appendix
\section{Full dataset} 

\begin {table*}[h!]
\begin{center}
\caption {Full available dataset. Radial velocities are given in the Barycentric Reference Frame after subtracting the secular acceleration. Radial-velocity uncertainties include photon noise, calibration and telescope related uncertainties.\\
 BJD is referenced to day 2450000, <V$_{r ~CCF}$> = 12850.11 ms$^{-1}$, <FWHM> 3703.09 ms$^{-1}$. \label{tab:full_data}}
    \begin{tabular}{ c  c  c c c c c c c c c c c} \hline
BJD & $\Delta$V$_{r ~CCF}$  & $\sigma$ V$_{r ~CCF}$ & $\Delta$V$_{r ~TERRA}$  & $\sigma$ V$_{r ~TERRA}$ &  $\Delta$FWHM  & BIS Span  & S$_{MW}$ & $\sigma$ S$_{MW}$ & H$_{\alpha}$ & $\sigma$ H$_{\alpha}$  \\ 
 (d) & (ms$^{-1}$) & (ms$^{-1}$) & (ms$^{-1}$) & (ms$^{-1}$) & (ms$^{-1}$) & (ms$^{-1}$) \\ \hline
6438.576	& 2.00 & 1.31& 2.64& 1.19 & -2.48 & -7.68 & 0.620 & 0.009 & 0.4260 & 0.0006 \\ 
6440.577	& -2.63 & 1.25& -0.22& 1.25 & 1.77 & -6.84 & 0.627 & 0.008 & 0.4164 & 0.0005 \\ 
6441.698	& -0.08 & 1.33& 1.63& 1.31 & -0.82 & -8.32 & 0.623 & 0.008 & 0.4210 & 0.0005 \\ 
6442.513	& -0.88 & 1.21& -0.83& 1.22 & 5.56 & -8.24 & 0.728 & 0.008 & 0.4300 & 0.0005 \\ 
6443.481	& -1.17 & 1.52& -0.33& 1.34 & -1.57 & -4.93 & 0.654 & 0.012 & 0.4157 & 0.0008 \\ 
6484.636	& -5.47 & 1.65& -4.33& 1.33 & 2.81 & -11.65 & 0.529 & 0.011 & 0.4124 & 0.0008 \\ 
6485.458	& -4.32 & 1.30& -4.40& 1.21 & -2.18 & -10.24 & 0.595 & 0.008 & 0.4137 & 0.0005 \\ 
6507.464	& 2.71 & 2.10& 2.98& 1.90 & -6.48 & -12.12 & 0.553 & 0.013 & 0.4069 & 0.0009 \\ 
6533.354	& -0.90 & 1.29& 0.67& 1.25 & 1.38 & -8.02 & 0.614 & 0.008 & 0.4187 & 0.0005 \\ 
6534.363	& -4.75 & 1.65& -2.98& 1.38 & -5.46 & -7.17 & 0.601 & 0.012 & 0.4221 & 0.0008 \\ 
6535.419	& -0.21 & 1.45& -0.49& 1.29 & -0.76 & -6.18 & 0.592 & 0.012 & 0.4141 & 0.0008 \\ 
6536.475	& -0.84 & 1.29& 1.26& 1.27 & -2.85 & -8.30 & 0.684 & 0.008 & 0.4143 & 0.0005 \\ 
6693.714	& -2.83 & 1.51& -1.58& 1.39 & 0.19 & -5.41 & 0.744 & 0.013 & 0.4199 & 0.0007 \\ 
6694.738	& -2.29 & 1.46& -1.86& 1.38 & 1.60 & -10.44 & 0.721 & 0.012 & 0.4175 & 0.0007 \\ 
6695.707	& 3.72 & 2.04& 5.27& 1.52 & -0.13 & -6.90 & 0.663 & 0.017 & 0.4140 & 0.0010 \\ 
6696.698	& -1.52 & 1.53& 0.97& 1.30 & 2.95 & -5.39 & 0.692 & 0.012 & 0.4170 & 0.0007 \\ 
6697.710	& 1.71 & 1.28& 3.68& 1.25 & 3.28 & -7.19 & 0.821 & 0.011 & 0.4309 & 0.0006 \\ 
6698.711	& 0.93 & 1.61& 2.22& 1.37 & 0.84 & -4.92 & 0.663 & 0.013 & 0.4203 & 0.0008 \\ 
6700.690	& -0.65 & 1.31& 0.92& 1.26 & 1.72 & -6.98 & 0.792 & 0.011 & 0.4158 & 0.0006 \\ 
6701.667	& 3.17 & 1.46& 4.74& 1.33 & -5.70 & -3.22 & 0.717 & 0.013 & 0.4161 & 0.0008 \\ 
6702.662	& -1.44 & 1.45& 1.79& 1.43 & -4.22 & -6.69 & 0.769 & 0.011 & 0.4278 & 0.0007 \\ 
6775.629	& 1.08 & 1.75& 2.53& 1.39 & 2.65 & -10.16 & 0.779 & 0.016 & 0.4069 & 0.0010 \\ 
6783.490	& 3.03 & 1.32& 3.46& 1.31 & 5.21 & -7.24 & 0.818 & 0.012 & 0.4229 & 0.0007 \\ 
6784.494	& 0.99 & 1.30& 1.62& 1.20 & 6.69 & -4.60 & 0.857 & 0.011 & 0.4262 & 0.0006 \\ 
6798.430	& -1.74 & 1.29& -0.87& 1.22 & 0.89 & -6.52 & 0.815 & 0.010 & 0.4233 & 0.0006 \\ 
6799.453	& 0.95 & 1.28& 1.36& 1.23 & -0.99 & -6.34 & 0.833 & 0.011 & 0.4266 & 0.0006 \\ 
6800.417	& 0.31 & 1.41& 3.22& 1.29 & 2.46 & -5.21 & 0.894 & 0.013 & 0.4371 & 0.0007 \\ 
6821.485	& -2.71 & 1.84& -3.05& 1.51 & -7.94 & -4.35 & 0.590 & 0.016 & 0.4217 & 0.0011 \\ 
6854.495	& -1.17 & 1.25& 0.21& 1.25 & -1.00 & -5.66 & 0.931 & 0.011 & 0.4235 & 0.0006 \\ 
6855.484	& -1.44 & 1.32& -0.18& 1.25 & -0.89 & -7.25 & 0.746 & 0.011 & 0.4091 & 0.0006 \\ 
6857.467	& -0.71 & 1.29& 0.15& 1.21 & 3.35 & -4.95 & 0.813 & 0.012 & 0.4049 & 0.0007 \\ 
6858.436	& -0.09 & 1.35& 1.39& 1.29 & 4.39 & -7.70 & 0.777 & 0.011 & 0.4093 & 0.0006 \\ 
6859.458	& 4.93 & 1.32& 4.60& 1.28 & 3.63 & -9.58 & 0.761 & 0.010 & 0.4075 & 0.0006 \\ 
6860.423	& 2.78 & 1.24& 3.02& 1.17 & 4.26 & -6.83 & 0.845 & 0.010 & 0.4127 & 0.0006 \\ 
6861.450	& 2.21 & 1.27& 4.23& 1.24 & 7.78 & -5.48 & 0.994 & 0.011 & 0.4216 & 0.0005 \\ 
6877.444	& 2.30 & 1.30& 3.25& 1.19 & 2.60 & -8.93 & 0.813 & 0.011 & 0.4292 & 0.0006 \\ 
6878.442	& 2.18 & 1.23& 2.98& 1.25 & 4.06 & -5.39 & 0.781 & 0.010 & 0.4154 & 0.0005 \\ 
6879.383	& -1.02 & 1.24& -0.29& 1.19 & 6.14 & -4.11 & 0.889 & 0.012 & 0.4259 & 0.0006 \\ 
6880.386	& 1.05 & 1.36& 2.13& 1.25 & 4.76 & -3.22 & 0.781 & 0.012 & 0.4214 & 0.0007 \\ 
6881.403	& 0.84 & 1.95& 0.81& 1.47 & -1.91 & -6.34 & 0.793 & 0.020 & 0.4276 & 0.0011 \\ 
6892.395	& 1.44 & 1.31& 2.62& 1.21 & -0.40 & -7.09 & 0.736 & 0.011 & 0.4132 & 0.0006 \\ 
6893.398	& 3.93 & 1.45& 2.50& 1.29 & 3.29 & -6.48 & 0.711 & 0.012 & 0.4140 & 0.0007 \\ 
6894.401	& 2.30 & 1.39& 1.50& 1.30 & 2.56 & -6.94 & 0.716 & 0.010 & 0.4088 & 0.0006 \\ 
6897.410	& -2.36 & 1.29& -1.97& 0.98 & -0.59 & -3.46 & 0.757 & 0.015 & 0.4259 & 0.0009 \\ 
6898.381	& -4.20 & 1.39& -1.51& 1.07 & -1.21 & -7.39 & 0.791 & 0.015 & 0.4251 & 0.0009 \\ 
6899.395	& -0.69 & 1.26& 0.34& 0.98 & -4.95 & -0.65 & 0.734 & 0.017 & 0.4186 & 0.0010 \\ 
6904.495	& -2.25 & 1.01& -0.90& 0.90 & -2.00 & -6.76 & 0.760 & 0.009 & 0.4263 & 0.0006 \\ 
6905.388	& 1.03 & 0.99& 1.21& 0.94 & -6.51 & -4.09 & 0.767 & 0.011 & 0.4209 & 0.0006 \\ 
6907.382	& -2.05 & 1.02& -0.46& 0.92 & -10.51 & -5.15 & 0.694 & 0.010 & 0.4182 & 0.0006 \\ 
6909.380	& -4.11 & 1.23& -2.25& 1.24 & -7.44 & -4.52 & 0.707 & 0.009 & 0.4196 & 0.0006 \\ 
6918.389	& -3.37 & 1.48& -3.10& 1.18 & -8.96 & -7.26 & 0.595 & 0.013 & 0.4138 & 0.0009 \\ 
6920.388	& -2.44 & 1.09& -1.45& 0.93 & -13.49 & -10.08 & 0.670 & 0.010 & 0.4172 & 0.0007 \\ 
 6921.393	& -1.43 & 1.78& -1.85& 1.38 & -15.38 & -6.81 & 0.658 & 0.016 & 0.4167 & 0.0010 \\ 
 6932.342	& -1.54 & 1.45& -1.60& 1.26 & -5.48 & -9.18 & 0.706 & 0.016 & 0.4190 & 0.0010 \\ 
 6938.320	& -1.49 & 1.08& 0.61& 0.97 & -1.03 & -7.88 & 0.779 & 0.012 & 0.4142 & 0.0007 \\ 
 6940.320	& -0.47 & 1.16& -0.94& 1.36 & -0.26 & -5.09 & 0.830 & 0.013 & 0.4176 & 0.0007 \\ 
 6942.318	& 1.43 & 0.93& 1.41& 0.90 & 4.59 & -6.06 & 0.909 & 0.010 & 0.4208 & 0.0005 \\ 

\hline
\label{full_data}
\end{tabular}  
\end{center}
\end {table*}

\begin {table*}
\ContinuedFloat
\begin{center}
\caption {Full available dataset. Radial velocities are given in the Barycentric Reference Frame after subtracting the secular acceleration. Radial-velocity uncertainties include photon noise, calibration and telescope related uncertainties.\\
 BJD is referenced to day 2450000, <V$_{r ~CCF}$> = 12850.11 ms$^{-1}$, <FWHM> 3703.09 ms$^{-1}$. \label{tab:full_data}}
    \begin{tabular}{ c  c  c c c c c c c c c c c} \hline
BJD & $\Delta$V$_{r ~CCF}$  & $\sigma$ V$_{r ~CCF}$ & $\Delta$V$_{r ~TERRA}$  & $\sigma$ V$_{r ~TERRA}$ &  $\Delta$FWHM  & BIS Span  & S$_{MW}$ & $\sigma$ S$_{MW}$ & H$_{\alpha}$ & $\sigma$ H$_{\alpha}$  \\ 
 (d) & (ms$^{-1}$) & (ms$^{-1}$) & (ms$^{-1}$) & (ms$^{-1}$) & (ms$^{-1}$) & (ms$^{-1}$) \\ \hline

6943.315	& 1.45 & 1.07& 0.40& 0.95 & 4.47 & -5.65 & 0.831 & 0.012 & 0.4140 & 0.0006 \\  
7113.565	& -3.73 & 1.63& -2.42& 1.41 & -7.34 & -2.10 & 0.773 & 0.017 & 0.4062 & 0.0009 \\ 
7114.622	& -5.75 & 3.47& -3.72& 2.51 & -4.94 & -10.36 & 0.769 & 0.034 & 0.4065 & 0.0019 \\ 
7115.529	& -2.29 & 1.58& -2.15& 1.30 & -9.84 & -5.27 & 0.629 & 0.013 & 0.4031 & 0.0008 \\ 
7116.576	& -2.83 & 1.26& -1.73& 1.18 & -4.91 & -5.72 & 0.660 & 0.009 & 0.4038 & 0.0005 \\ 
7123.548	& 2.42 & 3.67& 1.08& 2.45 & 0.81 & -6.15 & 0.607 & 0.032 & 0.4107 & 0.0022 \\ 
7137.713	& 2.10 & 1.36& 5.33& 1.35 & -6.38 & -7.25 & 0.676 & 0.009 & 0.4073 & 0.0005 \\ 
7139.674	& 4.80 & 1.44& 3.90& 1.29 & -3.61 & -5.28 & 0.650 & 0.011 & 0.4055 & 0.0007 \\ 
7142.604	& -0.34 & 1.36& 0.45& 1.25 & -2.94 & -3.25 & 0.720 & 0.012 & 0.4117 & 0.0007 \\ 
7143.549	& -1.32 & 1.53& -2.04& 1.35 & -1.85 & -6.11 & 0.615 & 0.013 & 0.4071 & 0.0008 \\ 
7144.543	& 0.95 & 1.52& 1.87& 1.35 & -5.68 & -7.39 & 0.760 & 0.015 & 0.4243 & 0.0009 \\ 
7239.410	& 1.79 & 1.01& 0.90& 0.95 & -1.23 & -7.22 & 0.761 & 0.010 & 0.4157 & 0.0006 \\ 
7240.410	& 5.76 & 1.03& 6.15& 0.95 & 0.21 & -4.99 & 0.726 & 0.011 & 0.4135 & 0.0006 \\ 
7241.407	& 0.71 & 1.30& 2.84& 1.09 & -5.87 & -9.06 & 0.723 & 0.013 & 0.4145 & 0.0008 \\ 
7242.407	& 2.13 & 1.28& 3.25& 1.05 & -4.76 & -4.11 & 0.755 & 0.014 & 0.4269 & 0.0008 \\ 
7249.438	& -1.31 & 1.35& -2.59& 1.19 & 1.21 & -5.66 & 0.742 & 0.015 & 0.4135 & 0.0009 \\ 
7251.395	& -0.40 & 0.99& -1.04& 0.87 & 2.22 & -5.80 & 0.798 & 0.011 & 0.4209 & 0.0006 \\ 
7260.370	& -0.84 & 0.94& -0.84& 0.90 & -0.80 & -4.44 & 0.777 & 0.010 & 0.4146 & 0.0006 \\ 
7261.379	& 0.84 & 1.05& -0.59& 1.06 & 1.46 & -7.19 & 0.754 & 0.011 & 0.4099 & 0.0006 \\ 
7262.376	& -0.36 & 1.04& -0.47& 0.92 & 3.59 & -3.07 & 0.803 & 0.012 & 0.4212 & 0.0007 \\ 
7263.376	& 1.33 & 1.00& 1.77& 0.95 & 3.34 & -8.43 & 0.839 & 0.011 & 0.4176 & 0.0006 \\ 
7264.375	& -0.40 & 1.34& -0.34& 1.01 & 7.63 & -3.16 & 0.826 & 0.016 & 0.4247 & 0.0009 \\ 
7274.359	& 0.22 & 1.16& -0.10& 0.93 & 1.28 & -5.93 & 0.739 & 0.011 & 0.4018 & 0.0007 \\ 
7275.359	& -1.41 & 1.43& -1.71& 1.39 & 4.38 & -6.65 & 0.765 & 0.012 & 0.4069 & 0.0007 \\ 
7276.358	& -2.44 & 1.05& -3.20& 0.95 & 7.00 & -6.61 & 0.946 & 0.013 & 0.4176 & 0.0007 \\ 
7277.356	& -3.46 & 1.02& -3.28& 1.03 & 5.83 & -2.72 & 0.769 & 0.011 & 0.4045 & 0.0006 \\ 
7282.371	& 2.91 & 0.96& 5.24& 1.49 & 3.05 & -5.72 & 0.757 & 0.010 & 0.4013 & 0.0005 \\ 
7285.373	& 0.59 & 1.27& -0.99& 1.14 & 4.22 & -5.68 & 0.642 & 0.011 & 0.4093 & 0.0007 \\ 
7286.357	& -0.47 & 1.09& -0.74& 0.97 & -1.80 & -4.13 & 0.685 & 0.011 & 0.4050 & 0.0007 \\ 
7287.359	& 3.16 & 1.10& 3.46& 0.98 & 0.28 & -5.95 & 0.673 & 0.010 & 0.4070 & 0.0007 \\ 
7293.358	& 0.08 & 1.42& -1.26& 1.32 & 0.14 & -6.87 & 0.731 & 0.012 & 0.4110 & 0.0007 \\ 
7296.390	& 2.53 & 1.14& 3.24& 1.18 & -1.87 & -11.83 & 0.664 & 0.008 & 0.4146 & 0.0006 \\ 
7443.689	& 3.02 & 2.09& 2.59& 1.47 & -1.82 & -5.43 & 0.561 & 0.017 & 0.4027 & 0.0011 \\ 
7472.703	& 3.78 & 1.57& 4.04& 1.39 & -5.03 & -7.27 & 0.702 & 0.015 & 0.4161 & 0.0009 \\ 
7474.665	& 6.60 & 1.40& 5.81& 1.28 & -0.61 & -2.85 & 0.671 & 0.011 & 0.4117 & 0.0007 \\ 
7502.703	& -0.82 & 1.38& -0.55& 1.36 & -8.39 & -6.57 & 0.609 & 0.010 & 0.4148 & 0.0007 \\ 
7508.527	& -4.00 & 1.11& -1.33& 1.00 & -2.50 & -7.08 & 0.703 & 0.009 & 0.4204 & 0.0006 \\ 
7508.659	& -2.13 & 1.50& 0.54& 1.44 & -2.49 & -12.67 & 0.679 & 0.011 & 0.4170 & 0.0007 \\ 
7509.550	& -6.62 & 1.49& -7.26& 1.30 & -1.09 & -11.35 & 0.756 & 0.012 & 0.4202 & 0.0008 \\ 
7510.531	& -3.39 & 1.44& -3.98& 1.19 & -4.10 & -8.28 & 0.681 & 0.011 & 0.4141 & 0.0007 \\ 
7513.622	& -3.47 & 1.52& -4.33& 1.31 & -2.63 & -7.93 & 0.696 & 0.013 & 0.4148 & 0.0008 \\ 
7521.524	& -2.78 & 1.46& -1.79& 1.25 & -4.16 & -4.58 & 0.788 & 0.013 & 0.4094 & 0.0007 \\ 
7522.497	& 0.28 & 1.42& -0.78& 1.37 & 0.43 & -5.13 & 0.725 & 0.013 & 0.4071 & 0.0007 \\ 
7523.494	& -1.30 & 1.84& -1.64& 1.41 & 0.56 & -7.26 & 0.676 & 0.017 & 0.4083 & 0.0010 \\ 
7524.495	& -1.60 & 1.76& -0.93& 1.60 & 4.52 & -9.77 & 0.719 & 0.015 & 0.4090 & 0.0009 \\ 
7525.521	& -3.43 & 1.62& -3.72& 1.27 & 10.84 & -9.65 & 0.963 & 0.017 & 0.4281 & 0.0009 \\ 
7535.641	& -3.44 & 1.38& -4.24& 1.15 & 3.99 & -11.65 & 0.782 & 0.014 & 0.4101 & 0.0008 \\ 
7536.504	& -0.46 & 1.32& 1.46& 1.02 & 6.61 & -4.87 & 0.802 & 0.014 & 0.4090 & 0.0008 \\ 
7537.496	& -0.19 & 1.56& -0.35& 1.23 & 8.41 & -1.00 & 0.973 & 0.020 & 0.4306 & 0.0010 \\ 
7537.611	& 0.95 & 1.22& -1.19& 1.24 & 6.29 & -5.58 & 0.881 & 0.013 & 0.4135 & 0.0007 \\ 
7538.489	& -0.15 & 1.06& -0.73& 1.00 & 2.95 & -7.15 & 0.860 & 0.010 & 0.4218 & 0.0006 \\ 
7538.620	& -0.05 & 1.15& -0.16& 1.02 & 4.42 & -8.91 & 0.839 & 0.010 & 0.4101 & 0.0006 \\ 
7540.658	& -2.09 & 1.31& 0.11& 1.04 & 3.40 & -10.09 & 0.724 & 0.012 & 0.4062 & 0.0007 \\ 
7549.591	& 7.97 & 1.35& 8.69& 1.25 & -3.93 & -11.09 & 0.747 & 0.010 & 0.4118 & 0.0007 \\ 
7550.597	& 5.93 & 1.23& 6.33& 1.11 & 5.35 & -9.50 & 0.727 & 0.009 & 0.4206 & 0.0006 \\ 
7551.582	& 0.78 & 1.48& -0.33& 1.09 & -1.05 & -7.63 & 0.614 & 0.012 & 0.4151 & 0.0008 \\ 
7552.554	& -0.39 & 1.26& 0.62& 1.07 & -0.11 & -7.40 & 0.632 & 0.011 & 0.4142 & 0.0007 \\ 

 7553.561	& -2.19 & 1.43& -1.69& 1.21 & -0.30 & -6.43 & 0.655 & 0.013 & 0.4142 & 0.0009 \\ 
 \hline
\label{full_data}
\end{tabular}  
\end{center}
\end {table*}

\begin {table*}
\ContinuedFloat
\begin{center}
\caption {Full available dataset. Radial velocities are given in the Barycentric Reference Frame after subtracting the secular acceleration. Radial-velocity uncertainties include photon noise, calibration and telescope related uncertainties.\\
 BJD is referenced to day 2450000, <V$_{r ~CCF}$> = 12850.11 ms$^{-1}$, <FWHM> 3703.09 ms$^{-1}$. \label{tab:full_data}}
    \begin{tabular}{ c  c  c c c c c c c c c c c} \hline
BJD & $\Delta$V$_{r ~CCF}$  & $\sigma$ V$_{r ~CCF}$ & $\Delta$V$_{r ~TERRA}$  & $\sigma$ V$_{r ~TERRA}$ &  $\Delta$FWHM  & BIS Span  & S$_{MW}$ & $\sigma$ S$_{MW}$ & H$_{\alpha}$ & $\sigma$ H$_{\alpha}$  \\ 
 (d) & (ms$^{-1}$) & (ms$^{-1}$) & (ms$^{-1}$) & (ms$^{-1}$) & (ms$^{-1}$) & (ms$^{-1}$) \\ \hline

7594.464	& 0.13 & 1.51& -0.26& 1.35 & 1.00 & -5.21 & 0.817 & 0.012 & 0.4181 & 0.0007 \\
7596.458	& 0.77 & 1.52& -0.48& 1.45 & 1.47 & -5.79 & 0.761 & 0.014 & 0.4188 & 0.0008 \\ 
7597.470	& 1.36 & 1.53& 0.71& 1.43 & 1.71 & -10.35 & 0.702 & 0.012 & 0.4167 & 0.0007 \\ 
7603.456	& -0.63 & 1.36& 0.19& 1.32 & 7.67 & -8.98 & 0.757 & 0.009 & 0.4093 & 0.0006 \\ 
7604.446	& -0.13 & 1.61& -0.43& 1.43 & 9.40 & -6.31 & 0.779 & 0.014 & 0.4111 & 0.0008 \\ 
7606.393	& 4.30 & 1.47& 1.86& 1.19 & 10.04 & -5.10 & 0.765 & 0.014 & 0.4124 & 0.0008 \\ 
7607.474	& 5.68 & 2.20& 5.39& 2.23 & 16.25 & -7.92 & 0.763 & 0.018 & 0.4103 & 0.0011 \\ 
7608.435	& 0.46 & 1.12& -0.90& 1.02 & 9.23 & -6.53 & 0.802 & 0.011 & 0.4103 & 0.0006 \\ 
7609.430	& -1.96 & 3.14& -2.40& 2.19 & 13.38 & -7.73 & 0.817 & 0.030 & 0.4128 & 0.0018 \\ 
7610.400	& 0.15 & 2.67& 0.40& 2.15 & 8.52 & -7.85 & 0.983 & 0.031 & 0.4348 & 0.0016 \\ 
7620.430	& 1.97 & 1.68& 0.52& 1.45 & -0.40 & -6.87 & 0.668 & 0.015 & 0.4125 & 0.0009 \\ 
7621.461	& 1.71 & 2.81& -0.53& 1.80 & 11.63 & -12.30 & 0.436 & 0.018 & 0.4147 & 0.0015 \\ 
7625.405	& 0.43 & 1.02& -0.33& 1.10 & -4.17 & -4.94 & 0.815 & 0.011 & 0.4158 & 0.0006 \\ 
7626.409	& 0.61 & 1.38& 0.20& 1.18 & -5.01 & -6.14 & 0.710 & 0.014 & 0.4111 & 0.0009 \\ 
7627.403	& -1.73 & 1.22& 0.86& 1.06 & -6.58 & -4.51 & 0.687 & 0.013 & 0.4175 & 0.0008 \\ 
7628.406	& -1.25 & 1.21& -1.71& 1.13 & -5.40 & -3.69 & 0.814 & 0.011 & 0.4267 & 0.0007 \\ 
7629.402	& 2.80 & 1.15& 2.58& 1.01 & -0.81 & -7.01 & 0.700 & 0.012 & 0.4179 & 0.0007 \\ 
7632.404	& -0.48 & 1.09& -0.26& 0.98 & -6.03 & -6.18 & 0.744 & 0.011 & 0.4084 & 0.0006 \\ 
7637.400	& 3.40 & 1.15& 2.82& 1.09 & -5.64 & -7.31 & 0.653 & 0.010 & 0.4186 & 0.0007 \\ 
7638.385	& 1.80 & 1.16& 1.54& 1.12 & -5.09 & -8.60 & 0.676 & 0.012 & 0.4173 & 0.0008 \\ 
7641.383	& -0.89 & 1.00& -1.85& 1.02 & -5.71 & -7.20 & 0.650 & 0.009 & 0.4261 & 0.0006 \\ 
7642.382	& -2.00 & 1.07& -2.65& 1.03 & -3.46 & -4.29 & 0.627 & 0.010 & 0.4199 & 0.0007 \\ 
7643.380	& -1.43 & 1.15& -2.24& 1.09 & -4.58 & -5.93 & 0.689 & 0.012 & 0.4283 & 0.0007 \\ 
7645.381	& -1.98 & 1.18& -1.15& 1.16 & -7.10 & -8.70 & 0.640 & 0.011 & 0.4276 & 0.0007 \\ 
7646.377	& -1.07 & 1.14& -1.08& 0.98 & -5.02 & -7.60 & 0.670 & 0.011 & 0.4161 & 0.0007 \\ \hline

\label{full_data}
\end{tabular}  
\end{center}
\end {table*}

\begin {table}
\begin{center}
\caption {EXORAP B band photometric dataset \label{tab:full_data_B}}
    \begin{tabular}{ c  c  c } \hline
MJD & $\Delta$m$_{B}$  & $\sigma$ m$_{B}$  \\ 
 (d) & (mmag) & (mmag)  \\ \hline
6782.913	& 29.78 & 10.13 \\ 
6784.956	& 30.38 & 10.10 \\ 
6786.940	& 28.48 & 10.11 \\ 
6797.895	& 19.68 & 10.14 \\ 
6798.910	& 18.58 & 10.11 \\ 
6801.896	& 30.18 & 10.11 \\ 
6802.888	& 30.78 & 10.11 \\ 
6813.872	& 26.58 & 10.10 \\ 
6817.853	& 22.18 & 10.13 \\ 
6830.982	& 9.68 & 10.13 \\ 
6834.832	& 10.38 & 10.16 \\ 
6835.832	& 9.68 & 10.11 \\ 
6836.832	& 10.08 & 10.11 \\ 
6839.823	& 0.58 & 10.11 \\ 
6840.822	& 11.48 & 10.13 \\ 
6841.822	& 6.38 & 10.20 \\ 
6845.876	& 4.28 & 10.16 \\ 
6851.846	& 20.38 & 10.13 \\ 
6866.853	& 27.68 & 10.10 \\ 
6868.852	& 21.48 & 10.10 \\ 
7037.086	& -0.92 & 10.11 \\ 
7110.987	& 4.38 & 10.13 \\ 
7112.016	& -3.82 & 10.10 \\ 
7113.014	& 1.68 & 10.10 \\ 
7117.946	& 16.18 & 10.10 \\ 
7154.924	& 20.78 & 10.07 \\ 
7162.053	& 13.78 & 10.06 \\ 
7165.006	& 31.08 & 10.06 \\ 
7165.970	& 14.48 & 10.05 \\ 
7171.845	& 8.18 & 10.06 \\ 
7176.994	& -0.22 & 10.07 \\ 
7184.862	& 8.28 & 10.08 \\ 
7196.035	& 2.18 & 10.02 \\ 
7199.824	& 4.38 & 10.05 \\ 
7200.864	& 5.28 & 10.04 \\ 
7204.069	& -8.42 & 10.04 \\ 
7208.096	& -27.02 & 10.08 \\ 
7208.919	& -19.02 & 10.05 \\ 
7209.918	& -14.72 & 10.03 \\ 
7211.064	& -26.02 & 10.05 \\ 
7211.912	& -13.12 & 10.03 \\ 
7212.913	& -14.62 & 10.04 \\ 
7214.925	& -9.82 & 10.03 \\ 
7216.057	& -24.32 & 10.03 \\ 
7222.052	& -23.82 & 10.08 \\ 
7222.963	& -12.92 & 10.05 \\ 
7224.958	& -9.02 & 10.04 \\ 
7225.955	& -11.42 & 10.03 \\ 
7227.897	& -6.52 & 10.04 \\ 
7231.864	& 4.88 & 10.03 \\ 
7232.948	& 6.18 & 10.04 \\ 
7239.991	& -16.52 & 10.08 \\ 
7240.984	& -12.22 & 10.06 \\ 
7247.969	& 0.28 & 10.07 \\ 
7258.951	& 5.28 & 10.06 \\ 
7386.203	& 2.38 & 10.04 \\ 
7398.172	& -3.22 & 10.04 \\ 
7416.127	& -7.52 & 10.04 \\ 
7417.128	& -4.72 & 10.04 \\  \hline

\label{full_data_B}
\end{tabular}  
\end{center}
\end {table}

\begin {table}
\ContinuedFloat
\begin{center}
\caption {EXORAP B band photometric dataset (continued)\label{tab:full_data_B}}
    \begin{tabular}{ c  c  c } \hline
MJD & $\Delta$m$_{B}$  & $\sigma$ m$_{B}$  \\ 
 (d) & (mmag) & (mmag)  \\ \hline
7419.097	& -10.72 & 10.04 \\ 
7424.121	& -6.42 & 10.04 \\ 
7425.115	& -4.02 & 10.06 \\ 
7437.081	& -5.32 & 10.04 \\ 
7439.068	& -9.32 & 10.07 \\ 
7440.054	& -8.52 & 10.06 \\ 
7443.055	& 1.78 & 10.10 \\ 
7445.055	& 9.78 & 10.11 \\ 
7455.091	& 6.38 & 10.05 \\ 
7483.003	& 0.78 & 10.06 \\ 
7483.998	& 0.48 & 10.07 \\ 
7488.014	& 1.78 & 10.05 \\ 
7488.995	& 7.68 & 10.05 \\ 
7489.993	& 7.48 & 10.06 \\ 
7490.987	& -0.82 & 10.07 \\ 
7491.988	& 1.38 & 10.08 \\ 
7493.979	& 2.18 & 10.06 \\ 
7494.957	& -2.12 & 10.07 \\ 
7495.949	& 6.88 & 10.10 \\ 
7497.936	& -3.12 & 10.06 \\ 
7498.922	& -8.52 & 10.10 \\ 
7504.949	& -7.32 & 10.05 \\ 
7507.953	& -6.12 & 10.05 \\ 
7511.993	& -5.12 & 10.05 \\ 
7515.073	& -12.12 & 10.08 \\ 
7517.002	& -8.52 & 10.11 \\ 
7518.027	& -3.92 & 10.05 \\ 
7523.995	& -1.02 & 10.05 \\ 
7524.922	& 12.78 & 10.06 \\ 
7545.962	& -4.02 & 10.06 \\ 
7546.958	& 1.48 & 10.06 \\ 
7550.949	& -2.82 & 10.08 \\ 
7552.965	& -7.42 & 10.06 \\ 
7553.904	& 1.48 & 10.06 \\ 
7554.847	& -2.92 & 10.07 \\ 
7565.945	& -11.72 & 10.06 \\ 
7569.959	& -11.12 & 10.08 \\ 
7570.957	& -9.22 & 10.08 \\ 
7571.956	& -17.82 & 10.06 \\ 
7573.957	& -10.12 & 10.06 \\ 
7574.957	& -8.02 & 10.08 \\ 
7575.990	& -13.42 & 10.08 \\ 
7576.989	& -12.32 & 10.06 \\ 
7577.987	& -13.62 & 10.08 \\ 
7578.987	& -19.92 & 10.10 \\ 
7590.945	& -11.42 & 10.10 \\ 
7591.949	& -24.12 & 10.11 \\ 
7593.977	& -17.82 & 10.10 \\ 
7746.197	& -10.02 & 10.07 \\ 
 \hline
\label{full_data_B}
\end{tabular}  
\end{center}
\end {table}

\begin {table}
\begin{center}
\caption {EXORAP V band photometric dataset \label{tab:full_data_V}}
    \begin{tabular}{ c  c  c } \hline
MJD & $\Delta$m$_{V}$  & $\sigma$ m$_{V}$  \\ 
 (d) & (mmag) & (mmag)  \\ \hline
6782.914	& 25.43 & 10.05 \\ 
6784.957	& 21.23 & 10.04 \\ 
6785.950	& 22.23 & 10.08 \\ 
6786.941	& 40.13 & 10.05 \\ 
6787.862	& 26.33 & 10.06 \\ 
6797.896	& 10.63 & 10.06 \\ 
6798.912	& 29.03 & 10.05 \\ 
6801.897	& 14.23 & 10.05 \\ 
6802.889	& 10.63 & 10.05 \\ 
6812.873	& 14.03 & 10.04 \\ 
6816.832	& 4.43 & 10.05 \\ 
6830.983	& -4.37 & 10.05 \\ 
6832.849	& -4.67 & 10.07 \\ 
6834.833	& 4.23 & 10.06 \\ 
6835.833	& 1.03 & 10.05 \\ 
6836.833	& 1.83 & 10.05 \\ 
6839.824	& -10.77 & 10.05 \\ 
6840.824	& 2.23 & 10.06 \\ 
6841.824	& -3.47 & 10.07 \\ 
6842.823	& -11.57 & 10.08 \\ 
6845.877	& 0.23 & 10.07 \\ 
6851.848	& 8.83 & 10.05 \\ 
6863.041	& 2.73 & 10.06 \\ 
6866.854	& 14.63 & 10.05 \\ 
6868.853	& 16.33 & 10.05 \\ 
7037.087	& 3.63 & 10.05 \\ 
7112.017	& -6.57 & 10.04 \\ 
7113.015	& -6.97 & 10.04 \\ 
7117.947	& 2.93 & 10.04 \\ 
7154.925	& -7.87 & 10.03 \\ 
7165.008	& 0.93 & 10.02 \\ 
7184.863	& -21.77 & 10.03 \\ 
7196.036	& -1.07 & 10.02 \\ 
7199.825	& -0.07 & 10.06 \\ 
7200.865	& -1.97 & 10.02 \\ 
7204.070	& -8.67 & 10.02 \\ 
7208.097	& -16.47 & 10.03 \\ 
7208.920	& -12.27 & 10.02 \\ 
7211.065	& -15.87 & 10.02 \\ 
7211.913	& -15.87 & 10.02 \\ 
7222.054	& -14.87 & 10.04 \\ 
7222.964	& -15.07 & 10.02 \\ 
7231.865	& -3.27 & 10.02 \\ 
7239.992	& -1.67 & 10.05 \\ 
7240.985	& 1.63 & 10.04 \\ 
7247.970	& 1.43 & 10.04 \\ 
7258.952	& 5.03 & 10.04 \\ 
7386.204	& 6.73 & 10.03 \\ 
7398.173	& 9.33 & 10.03 \\ 
7416.128	& -0.77 & 10.03 \\ 
7417.129	& -3.87 & 10.03 \\ 
7419.098	& -3.57 & 10.03 \\ 
7424.122	& 4.43 & 10.03 \\ 
7425.116	& 2.43 & 10.05 \\ 
7437.082	& 1.93 & 10.03 \\ 
7439.069	& -5.77 & 10.05 \\ 
7440.055	& -1.57 & 10.04 \\ 
7445.056	& 7.43 & 10.03 \\
 \hline
\label{full_data_V}
\end{tabular}  
\end{center}
\end {table}

\begin {table}
\ContinuedFloat
\begin{center}
\caption {EXORAP V band photometric dataset (continued) \label{tab:full_data_V}}
    \begin{tabular}{ c  c  c } \hline
MJD & $\Delta$m$_{V}$  & $\sigma$ m$_{V}$  \\ 
 (d) & (mmag) & (mmag)  \\ \hline
7455.092	& 4.23 & 10.03 \\ 
7483.004	& -0.57 & 10.04 \\ 
7483.999	& 0.93 & 10.04 \\ 
7488.015	& 8.13 & 10.03 \\ 
7488.996	& 2.53 & 10.03 \\ 
7489.993	& -4.37 & 10.04 \\ 
7490.988	& -2.77 & 10.06 \\ 
7491.989	& -5.17 & 10.06 \\ 
7493.979	& 7.63 & 10.05 \\ 
7495.950	& 0.83 & 10.07 \\ 
7497.937	& -6.27 & 10.05 \\ 
7498.923	& -2.57 & 10.06 \\ 
7504.950	& -1.67 & 10.04 \\ 
7507.954	& -7.77 & 10.05 \\ 
7511.994	& -3.17 & 10.04 \\ 
7515.073	& -0.07 & 10.06 \\ 
7517.003	& -5.67 & 10.08 \\ 
7518.027	& 0.63 & 10.04 \\ 
7523.996	& 7.13 & 10.04 \\ 
7524.923	& 10.03 & 10.07 \\ 
7538.970	& 9.43 & 10.14 \\ 
7545.963	& -4.77 & 10.07 \\ 
7546.959	& 1.83 & 10.07 \\ 
7550.950	& -3.37 & 10.14 \\ 
7552.966	& -2.47 & 10.07 \\ 
7554.848	& -3.97 & 10.08 \\ 
7565.945	& -8.97 & 10.08 \\ 
7569.959	& -8.87 & 10.10 \\ 
7570.958	& -5.47 & 10.11 \\ 
7571.957	& -7.07 & 10.08 \\ 
7573.958	& -9.47 & 10.08 \\ 
7574.957	& -10.07 & 10.10 \\ 
7575.990	& -8.97 & 10.10 \\ 
7576.990	& -15.67 & 10.08 \\ 
7577.988	& -10.17 & 10.11 \\ 
7578.988	& -12.77 & 10.13 \\ 
7590.946	& -5.07 & 10.10 \\ 
7591.949	& -0.77 & 10.13 \\ 
7593.978	& -4.87 & 10.11 \\ 
7746.198	& -7.77 & 10.08 \\

 \hline
\label{full_data_V}
\end{tabular}  
\end{center}
\end {table}

\begin {table}
\begin{center}
\caption {EXORAP R band photometric dataset \label{tab:full_data_R}}
    \begin{tabular}{ c  c  c } \hline
MJD & $\Delta$m$_{V}$  & $\sigma$ m$_{V}$  \\ 
 (d) & (mmag) & (mmag)  \\ \hline
6782.915	& 36.87 & 10.05 \\ 
6784.957	& 42.47 & 10.04 \\ 
6785.951	& 34.37 & 10.08 \\ 
6786.941	& 44.07 & 10.04 \\ 
6787.863	& 45.67 & 10.06 \\ 
6798.912	& 45.77 & 10.05 \\ 
6802.890	& 30.27 & 10.05 \\ 
6830.984	& 21.57 & 10.05 \\ 
6832.849	& 17.57 & 10.07 \\ 
6834.834	& 20.47 & 10.08 \\ 
6835.834	& 18.17 & 10.05 \\ 
6839.825	& 17.57 & 10.05 \\ 
6841.824	& 12.17 & 10.08 \\ 
6845.878	& 22.17 & 10.07 \\ 
6851.848	& 27.07 & 10.05 \\ 
6863.042	& 9.17 & 10.06 \\ 
7037.088	& 1.67 & 10.05 \\ 
7112.017	& 3.17 & 10.03 \\ 
7113.016	& -11.23 & 10.04 \\ 
7117.947	& -8.23 & 10.03 \\ 
7154.925	& -4.53 & 10.03 \\ 
7165.008	& 5.67 & 10.02 \\ 
7184.863	& -8.83 & 10.03 \\ 
7196.037	& -3.13 & 10.02 \\ 
7199.826	& -5.83 & 10.03 \\ 
7200.866	& -1.93 & 10.02 \\ 
7204.071	& -19.03 & 10.02 \\ 
7208.098	& -34.83 & 10.02 \\ 
7211.065	& -17.73 & 10.02 \\ 
7222.054	& -29.33 & 10.02 \\ 
7231.866	& -5.13 & 10.02 \\ 
7239.992	& -5.83 & 10.05 \\ 
7240.985	& -10.53 & 10.04 \\ 
7247.970	& -12.63 & 10.05 \\ 
7258.953	& -1.73 & 10.05 \\ 
7386.204	& -0.33 & 10.04 \\ 
7398.173	& 0.17 & 10.04 \\ 
7416.129	& -14.53 & 10.04 \\ 
7417.129	& -10.33 & 10.04 \\ 
7419.098	& -8.53 & 10.04 \\ 
7424.123	& -9.73 & 10.04 \\ 
7425.117	& -9.03 & 10.06 \\ 
7437.082	& -0.53 & 10.04 \\ 
7439.069	& -4.03 & 10.05 \\ 
7440.056	& -3.03 & 10.05 \\ 
7445.056	& 1.17 & 10.04 \\ 
7455.093	& -4.13 & 10.04 \\ 
7483.004	& -2.13 & 10.05 \\ 
7483.999	& -1.23 & 10.06 \\ 
7488.016	& 12.77 & 10.04 \\ 
7488.996	& -7.33 & 10.04 \\ 
7489.994	& -3.53 & 10.05 \\ 
7490.989	& -5.03 & 10.06 \\ 
7491.990	& -7.73 & 10.06 \\ 
7493.980	& -7.13 & 10.05 \\ 
7495.950	& -7.33 & 10.06 \\ 
7497.938	& -8.43 & 10.05 \\ 
7498.923	& -15.43 & 10.05 \\ 
7504.950	& -6.03 & 10.04 \\ 
7507.955	& -9.93 & 10.04 \\ 
 \hline
\label{full_data_R}
\end{tabular}  
\end{center}
\end {table}
\label{lastpage}

\begin {table}
\ContinuedFloat
\begin{center}
\caption {EXORAP R band photometric dataset (continued) \label{tab:full_data_R}}
    \begin{tabular}{ c  c  c } \hline
MJD & $\Delta$m$_{V}$  & $\sigma$ m$_{V}$  \\ 
 (d) & (mmag) & (mmag)  \\ \hline
7511.994	& -8.33 & 10.04 \\ 
7515.074	& -17.83 & 10.06 \\ 
7517.003	& -14.53 & 10.08 \\ 
7518.028	& -5.63 & 10.04 \\ 
7524.923	& 10.47 & 10.08 \\ 
7538.970	& -4.53 & 10.13 \\ 
7545.963	& -3.13 & 10.10 \\ 
7546.959	& 0.67 & 10.10 \\ 
7552.966	& 0.87 & 10.10 \\ 
7554.848	& 6.37 & 10.13 \\ 
7565.946	& 4.57 & 10.11 \\ 
7569.960	& -13.13 & 10.14 \\ 
7570.958	& -5.33 & 10.14 \\ 
7571.957	& -18.33 & 10.11 \\ 
7573.958	& -7.23 & 10.11 \\ 
7574.958	& -14.33 & 10.14 \\ 
7575.991	& -22.83 & 10.14 \\ 
7576.990	& -6.63 & 10.11 \\ 
7578.988	& 0.97 & 10.16 \\ 
7590.946	& -4.83 & 10.13 \\ 
7593.978	& -13.23 & 10.14 \\ 
7746.198	& -8.33 & 10.08 \\

 \hline
\label{full_data_R}
\end{tabular}  
\end{center}
\end {table}

\begin {table}
\begin{center}
\caption {EXORAP I band photometric dataset \label{tab:full_data_I}}
    \begin{tabular}{ c  c  c } \hline
MJD & $\Delta$m$_{V}$  & $\sigma$ m$_{V}$  \\ 
 (d) & (mmag) & (mmag)  \\ \hline
6782.915	& 35.96 & 10.04 \\ 
6784.958	& 40.26 & 10.04 \\ 
6785.951	& 37.16 & 10.05 \\ 
6787.863	& 38.66 & 10.05 \\ 
6798.913	& 58.66 & 10.04 \\ 
6830.984	& 31.76 & 10.04 \\ 
6832.850	& 24.86 & 10.07 \\ 
6835.834	& 23.76 & 10.04 \\ 
6839.825	& 19.06 & 10.04 \\ 
6841.825	& 20.46 & 10.06 \\ 
6845.878	& 29.56 & 10.06 \\ 
6851.849	& 38.66 & 10.04 \\ 
6868.854	& 34.26 & 10.04 \\ 
7031.104	& -17.34 & 10.04 \\ 
7037.088	& -23.14 & 10.03 \\ 
7112.018	& -25.54 & 10.03 \\ 
7113.016	& -21.34 & 10.03 \\ 
7117.948	& -12.24 & 10.02 \\ 
7154.926	& -24.94 & 10.02 \\ 
7165.009	& -12.94 & 10.02 \\ 
7196.037	& -14.74 & 10.01 \\ 
7199.826	& -11.94 & 10.03 \\ 
7231.866	& -11.04 & 10.01 \\ 
7239.992	& 13.56 & 10.05 \\ 
7240.985	& -5.44 & 10.03 \\ 
7247.971	& 16.06 & 10.04 \\ 
7386.204	& -14.34 & 10.02 \\ 
7398.173	& -4.94 & 10.03 \\ 
7416.129	& -1.74 & 10.03 \\ 
7417.129	& -25.14 & 10.03 \\ 
7419.099	& 3.26 & 10.03 \\ 
7424.123	& -21.94 & 10.03 \\ 
7425.117	& -11.64 & 10.04 \\ 
7437.082	& -14.94 & 10.03 \\ 
7439.069	& -12.54 & 10.03 \\ 
7445.056	& 4.66 & 10.05 \\ 
7455.093	& -12.94 & 10.05 \\ 
7483.004	& -2.44 & 10.06 \\ 
7484.000	& -9.24 & 10.07 \\ 
7488.016	& -5.04 & 10.05 \\ 
7488.997	& -12.64 & 10.05 \\ 
7489.994	& -12.94 & 10.06 \\ 
7490.989	& -7.14 & 10.07 \\ 
7491.990	& -2.24 & 10.08 \\ 
7493.980	& -17.04 & 10.06 \\ 
7495.950	& -3.34 & 10.07 \\ 
7497.938	& -17.24 & 10.06 \\ 
7498.923	& -20.04 & 10.06 \\ 
7504.951	& -18.94 & 10.05 \\ 
7507.955	& -13.44 & 10.05 \\ 
7511.994	& -8.34 & 10.05 \\ 
7515.074	& -13.34 & 10.07 \\ 
7517.004	& -2.94 & 10.10 \\ 
7518.028	& -9.44 & 10.06 \\ 
7523.996	& 6.16 & 10.05 \\ 

 \hline
\label{full_data_I}
\end{tabular}  
\end{center}
\end {table}

\label{lastpage}

\end{document}